\documentclass[12pt]{article}
\usepackage[margin=1in, paperwidth=8.5in, paperheight=11in]{geometry}
\usepackage{amsmath}
\usepackage{graphicx}%
\usepackage{amsfonts}%
\usepackage{amssymb}
\usepackage{color, soul}
\usepackage{float,subfigure}
\usepackage{setspace}
\usepackage{comment}
\usepackage{setspace}     
\usepackage{chngpage}  
\usepackage{natbib}  
\usepackage{epsfig}
\usepackage{subfigure}
\usepackage{hyperref}
\usepackage{multirow}
\usepackage{ifsym}
\usepackage{caption}
\usepackage{wrapfig}
\usepackage{appendix}
\usepackage{layout}
\usepackage{titlesec}
\usepackage{rotating}
\usepackage{array,arydshln} 
\usepackage{pdflscape}
\usepackage{mathtools}  

\newcommand{\bbeta}{ \mbox{\boldmath $ \beta $} }

\newcommand{\balpha}{ \mbox{\boldmath $ \alpha $} }

\newcommand{\be}{ {\bf e} }

\newcommand{\bX}{ {\bf X} }

\newcommand{\bW}{ {\bf W} }

\newcommand{\given}{\,|\,}
\newcommand{\indep}{\rotatebox[origin=c]{90}{$\models$}}

\graphicspath{{figures/}}
\begin{document}
\title{Propensity score-based estimators with multiple error-prone covariates}
\author{Hwanhee Hong\thanks{Department of Mental Health, Bloomberg School of Public Health, Johns Hopkins University, Baltimore, Maryland, U.S.A.; {\tt hhong@jhu.edu}},
  David A. Aaby\thanks{Department of Psychiatry and Behavioral Sciences, Feinberg School of Medicine, Northwestern University, Chicago, Illinois, U.S.A.; {\tt david.aaby@northwestern.edu}},
  Juned Siddique\thanks{Department of Preventive Medicine, Feinberg School of Medicine, Northwestern University, Chicago, Illinois, U.S.A.; {\tt siddique@northwestern.edu}},
  Elizabeth A. Stuart\thanks{Department of Mental Health, Bloomberg School of Public Health, Johns Hopkins University, Baltimore, Maryland, U.S.A.; {\tt estuart@jhu.edu}}}
\date{\today}
\maketitle

\begin{abstract}
{Propensity score methods are an important tool to help reduce confounding in non-experimental studies. Most propensity score methods assume that covariates are measured without error. However, covariates are often measured with error, which leads to biased causal effect estimates if the true underlying covariates are the actual confounders. Although some studies have investigated the impact of a single mismeasured covariate on estimating a causal effect and proposed methods for handling the measurement error, almost no work exists investigating the case where multiple covariates are mismeasured. 
In this paper, we examine the consequences of multiple error-prone covariates when estimating causal effects using propensity score-based estimators via extensive simulation studies and real data analyses. We find that causal effect estimates are less biased when the propensity score model includes mismeasured covariates whose true underlying values are strongly correlated with each other. However, when the measurement \emph{errors} are correlated with each other, additional bias is introduced. 
In addition, it is beneficial to include correctly measured auxiliary variables that are correlated with confounders whose true underlying values are mismeasured in the propensity score model. \\
Running head: Propensity score estimators with error-prone covariates}
\end{abstract}

\section{Introduction}

In non-randomized epidemiological studies aiming to estimate causal effects, controlling for all potential confounders or covariates is essential to estimate unbiased causal effects. For this, propensity score methods are widely used to balance treatment (or exposure) groups in terms of observed covariates~\citep{Rosenbaum83}. The balance can be achieved properly when we collect all potential confounders and they are measured correctly. If these confounders are measured with error, the resulting estimates may be biased because we cannot directly balance the two groups with respect to the unobserved true confounders~\citep{Steiner11}. 

In practice, there is often more than one covariate measured with error. Measurement error in multiple covariates (vs. a single covariate) may yield additional complications when estimating causal effects, especially when the measurement errors themselves are correlated. This measurement error structure can be found in many applications. For example, it is known that self-reported test scores or mental health-related measures may not reflect the true underlying intellectual ability or psychometric profile. It is common that multiple self-reported scores are collected, and they tend to correlate with each other because they are obtained from the same individuals and using the same data collection strategy (e.g., in an in-person interview individuals may tend to systematically report fewer stigmatized behaviors than they actually engage in). In this example, if the unobserved true underlying covariates affect treatment assignment and an outcome (i.e., are confounders), using the self-reported measurements will not remove bias in the effect estimates because they cannot balance treatment groups with respect to the true confounders.     

Many propensity score approaches have been developed to handle a single error-prone covariate. Most approaches consider a classical measurement error case where a mismeasured covariate is a noisy yet unbiased version of a true covariate. These include propensity score calibration~\citep{Sturmer05}, corrected propensity score weighting~\citep{McCaffrey13a}, and calibrated propensity scores using multiple imputation~\citep{Webb-Vargas15}. A few approaches have addressed the situation where the measurement error depends on the treatment assignment or another covariate, called differential measurement error~\citep{Hong16}. We found one study investigating the impact of multiple error-prone covariates on a logistic regression model~\citep{Fewell07}. However, that study does not focus on a full causal framework including a step for balancing treatment groups with respect to observed covariates (such as through propensity scores). 

In this paper, we assess the impact of multiple error-prone covariates on causal effect estimation, especially when the measurement errors are correlated, using two widely-used propensity score estimators, inverse probability of treatment weighting (IPTW) and double robust (DR) estimators. In addition, we investigate the influence of measurement error in different types of covariates (e.g., covariates associated with the treatment assignment, outcome, or both) on causal effect estimation. The remainder of this paper is structured as follows. Section~\ref{sec:Background} overviews a general causal framework and the two propensity score-based causal effect estimators. Then, Section~\ref{sec:sim} provides settings and results of extensive simulation studies, and Section~\ref{sec:data} illustrates the impact of error-prone covariates using real data, the Observing Protein and Energy Nutrition (OPEN) Study by \cite{Subar03}. 
Finally, Section~\ref{sec:Discussion} discusses our work and needed future methodological developments.

\section{Background and setting} \label{sec:Background}

In this paper, we consider a binary treatment assignment $A$ ($A = 0$ or 1 for untreated or treated, respectively), a continuous outcome $Y$, continuous true covariates $\bX = (X_1, \dotso, X_p)$. We consider the scenario where $\bX$ is not observed but instead we observe $\bW = (W_1, \dotso, W_p)$ which is the mismeasurement of $\bX$. We assume classical measurement error where $W_j = X_j + e_j$ and $e_j {\sim} N(0, \sigma_{wj}^2)$, for $j=1, \dotso, p$. 

\subsection{Causal inference framework and assumptions}

The research question of interest is to estimate an effect of a treatment ($A=1$) as compared to a comparison condition ($A=0$) on a certain outcome. Our estimand of interest is the average treatment effect (ATE). Following the Rubin causal model~\citep{Rubin74}, the ATE is defined as $E(Y(1)-Y(0))$, where $Y(A = a)$ is the potential outcome under treatment assignment to $a$ (0 or 1). However, it is not feasible to observe both potential outcomes for an individual. In this paper, we are interested in non-experimental studies, where we simply observe that some people received the treatment and others received the control condition, and the treatment is not randomly assigned.     

In non-experimental studies, if all covariates ($\bX$) are correctly measured, the following assumptions are required to identify the ATE. First, there should be no unmeasured confounders, called \emph{ignorablility}, $Y(A=a) \indep A \given \bX $. Second, the potential outcome under an assigned treatment is consistent with the corresponding observed outcome, called \emph{consistency}, $Y(a) = (Y \given A=a)$. Third, there is a non-zero probability of receiving every level of treatment for every combination of values of covariates, called \emph{positivity}, $0 < P(A=a \given \bX) < 1$. An additional assumption is the stable unit treatment value assumption (SUTVA) meaning that one person's treatment assignment does not influence another person's potential outcomes and there is only one \emph{version} of each treatment~\citep{Rubin80}. 

\subsection{Propensity score weighting estimators}

To estimate an unbiased causal effect, we should balance treated and untreated groups with respect to the distribution of covariates. Propensity score methods are widely used to calculate balancing scores, functions of which are then sometimes used as weights~\citep{Rosenbaum83}. The propensity score, defined as $e_i(\bX_i) = P(A_i = 1 \given \bX_i)$ for subject $i$, is considered a summary score of all observed covariates and helps make the treated and control groups similar by matching, weighting, or subclassification instead of using the full set of covariates~\citep{Stuart10}. This is usually estimated by fitting a logistic (or probit) regression model. Propensity scores have two key properties that make them useful for causal effect estimation. First, propensity scores balance the treated and control groups with respect to the distribution of all \emph{observed} covariates, such as $\bX_i \indep A_i \given e_i(\bX_i)$. Second, if ignorability holds given $\bX_i$ then it also holds given $e_i(\bX_i)$, that is if $Y_i(a) \indep A_i \given \bX_i$ then $Y_i(a) \indep A_i \given e_i(X_i)$. 

The propensity score is a straightforward tool to handle non-random treatment assignment, but it is important to measure covariates correctly when they are associated with both treatment assignment and outcome. If we use mismeasured covariates $\bW$ instead of $\bX$ when estimating propensity scores, they cannot correctly balance the two groups with respect to $\bX$ and we would expect a biased ATE estimate. On the other hand, if $\bW$ is actually the variable associated with both treatment assignment and outcome (e.g., if a teacher used observed test scores to decide on program participation), using $\bW$ would not yield a biased estimate because ignorability would hold given $\bW$, not $\bX$, in this case.     

In this paper, we consider two propensity score weighting estimators: IPTW and DR. We first estimate propensity score $e_i$ by fitting a logistic regression of $A$ with covariates as predictors, and then calculate individual weights $w_i = A(1/e_i) + (1-A)(1/(1-e_i))$. The IPTW estimator is a generalization of the Horvitz-Thompson estimator~\citep{Horvitz52} and defined as    
\begin{equation} \notag
\widehat{ATE}_{IPTW} = \frac{\displaystyle\sum_i w_{i} A_i Y_i}{\displaystyle\sum_i w_{i} A_i} - \frac{\displaystyle\sum_i w_{i} (1-A_i) Y_i}{\displaystyle\sum_i w_{i} (1-A_i)}.
\end{equation}
To obtain $\widehat{ATE}_{IPTW}$ and its standard error, we fit a weighted linear regression model of $Y$ with $A$ as the only predictor using $w_{i}$ as weights. The coefficient for $A$ is taken as the effect estimate. We run this outcome model via the \texttt{survey} package in R. 

Doubly robust approaches can be used when the propensity score model is likely to be misspecified~\citep{Robins07}. These approaches are alternatives of IPTW because IPTW provides an unbiased estimate only when the propensity score model is correctly specified, while DR provides an unbiased effect estimate when either the propensity score or outcome model is correctly specified. The DR estimator is defined as
\begin{equation} \notag
\widehat{ATE}_{DR} = n^{-1} \sum_i[ w_{i} A_i Y_i - w_{i}(A_i - e_i)m_{1i}] - n^{-1} \sum_i[ w_{i} (1-A_i) Y_i + w_{i}(A_i - e_i)m_{0i}],
\end{equation}
where $m_{1i}$ and $m_{0i}$ are the predicted outcome values from the fitted outcome models $E(Y \given Z=1, \bX)$ and $E(Y \given Z=0, \bX)$, respectively. Following~\cite{McCaffrey13b}, we obtain a DR estimate similarly for an IPTW estimate via the \texttt{survey} package in R as described above, but we fit an outcome model (the weighted linear regression) including $A$ and all important covariates associated with the outcome as well.     

\section{Simulation studies} \label{sec:sim}

Our simulation studies are motivated by those in~\cite{Pingel16}, which studied the efficiency of different propensity score estimators when correlated covariates are observed. We extend the simulation setup to consider measurement error. We consider the implications for methods' performance of two types of correlations: first, when the true underlying covariates ($\bX$) are correlated, and second, when both the underlying covariates and their measurement errors ($\be$) are correlated. 
Our second simulation study investigates the consequences of measurement error in different types of covariates, such as covariates related to only the outcome or the treatment assignment, or both. In both simulation studies, we assess bias, mean squared error (MSE), and coverage probability of IPTW and DR estimates and compare these properties between when using true covariates ($\bX$) and mismeasured covariates ($\bW$).   

\subsection{Simulation 1: the impact of correlated covariates and measurement error}
\subsubsection{Data generating mechanism and parameter setup}

\begin{figure}[htp]
    \begin{center}
        \subfigure[Data structure for Simulation 1]{\includegraphics[height=5cm]{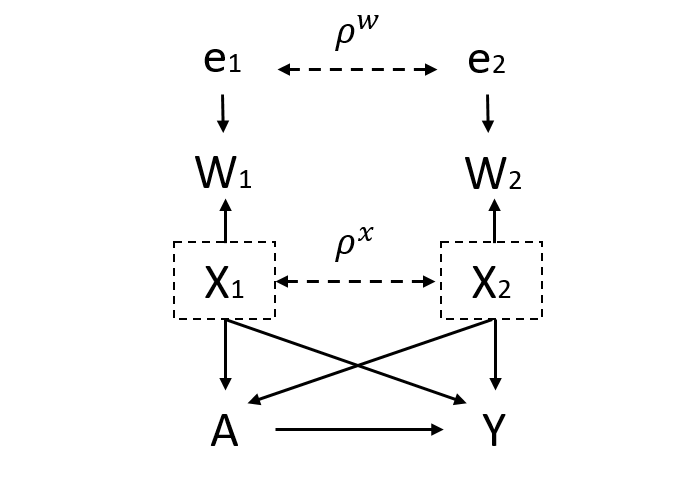}}
        \subfigure[Data structure for Simulation 2]{\includegraphics[height=5cm]{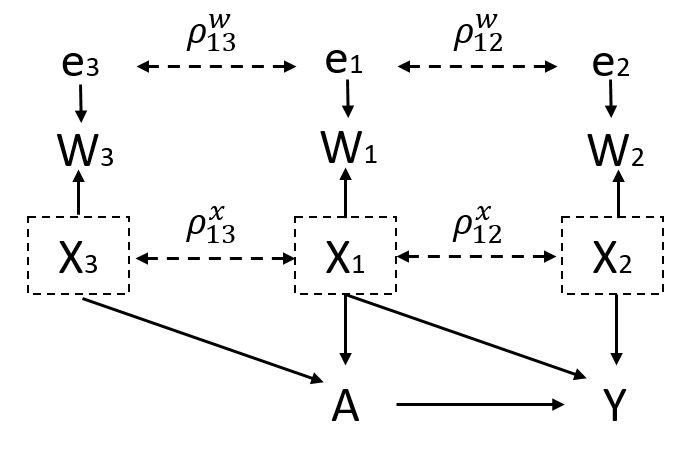}}
    \end{center}
\caption{Data generating mechanism for simulation studies. $X_p$ is the true covariates that are usually not observed in practice. $W_p$ is assumed to be $X_p + e_p$.}
\label{fig:simDAG}
\end{figure}

Panel (a) of Figure~\ref{fig:simDAG} illustrates the data structure we use in Simulation 1. We consider two confounders ($X_1$ and $X_2$) that are correlated with correlation $\rho^x$ and that are associated with both the treatment assignment and outcome. However, we observe only mismeasured covariates ($W_1$ and $W_2$) and the measurement errors are correlated (with correlation $\rho^w$). We generate 1000 sets of data where each dataset has 1000 observations under the following distributions:
\begin{align} \notag
\begin{pmatrix} X_1 \\ X_2 \end{pmatrix} {\sim} N\left( \begin{pmatrix} 0 \\ 0 \end{pmatrix}, \begin{pmatrix} 1 & \rho^x \\ \rho^x & 1 \end{pmatrix}  \right) \\ \label{eq:sim1_PS}
logit(P(A=1 \given \bX)) = \alpha_0 + \alpha_1 X_1 + \alpha_2 X_2 \\ \label{eq:sim1_Y}
Y \given A, \bX {\sim} N(\beta_0 + \tau A + \beta_1 X_1 + \beta_2 X_2, 1).
\end{align}
Then, we generate $W_p = X_p + e_p$ for $p=1, 2$, where 
\begin{equation} \notag
\begin{pmatrix} e_1 \\ e_2 \end{pmatrix} {\sim} N\left( \begin{pmatrix} 0 \\ 0 \end{pmatrix}, \begin{pmatrix} \sigma_{w_1}^2 & \sigma_{w_1}\sigma_{w_2}\rho^w \\ \sigma_{w_1}\sigma_{w_2}\rho^w & \sigma_{w_2}^2 \end{pmatrix}  \right). 
\end{equation}
We consider classical measurement error where the measurement errors are pure noise and not associated with the treatment assignment or the outcome. We define the reliability of a mismeasured covariate as $Var(X_p)/Var(W_p) = 1/(1+\sigma_{w_p}^2)$, where a smaller reliability value indicates a noisier $W_p$.  
For the true data generating models, we consider four different values for $\rho^x$ and $\rho^w$: $\rho^x = 0.0, 0.3, 0.6,$ or 0.9 and $\rho^w = 0.0, 0.2, 0.5,$ or 0.8. We set $\sigma_{w_p}^2$ to be 1, 0.43, or 0.1 resulting in corresponding reliabilities of 0.5, 0.7, and 0.9, respectively. 

We set the true coefficients, $\balpha$, $\tau$, and $\bbeta$, where $\balpha=(\alpha_0, \alpha_1, \alpha_2)^T$ and $\bbeta=(\beta_0, \beta_1, \beta_2)^T$, in Equations~\eqref{eq:sim1_PS} and~\eqref{eq:sim1_Y} in order to make simulated datasets under all different scenarios with different $\rho^x$ settings have the same amount of \emph{total confounding}. The total confounding is defined as the bias of a treatment effect estimate when excluding covariates that should be controlled for. As our simulation settings assume a linear relationship between the covariates and the outcome, the total confounding (see Appendix 1 for the derivation) is calculated as
\begin{equation} \label{eq:TC}
\sum_{p=1}^2 \beta_p [ E(X_p \given A=1) - E(X_p \given A=0) ]. 
\end{equation}
Here, given $Var(X_p) = 1$, $E(X_p \given A=1) - E(X_p \given A=0)$ is the standardized mean difference, a metric measuring a covariate imbalance between treatment groups. In addition, this is a function of $\rho^x$ so that the imbalance increases as $\rho^x$ increases given fixed positive $\balpha$ values. Note that the total confounding is not a function of $\rho^x$. When the total confounding is large we would naturally expect a large bias in an ATE estimate. Therefore, we need to set the true $\balpha$ and $\bbeta$ values carefully so that all simulated data under different settings contain the same amount of total confounding; otherwise differences in performance due to different values of $\rho^x$ may be partially due to different amounts of total confounding rather than due to $\rho^x$ itself. Thus, controlling the total confounding helps us more accurately assess the impact of correlation between true covariates on the ATE estimates when using mismeasured covariates. Table~\ref{table:Case1} shows the set up for $\balpha$, $\tau$, and $\bbeta$. We fix $\tau=2$, that is the true ATE is 2, and $\bbeta = (0, 1, 1)^T$. Then we select $\balpha$ values providing the total confounding closest to 1 for different $\rho^x$. These parameter setups result in approximately half the observations being assigned to the treated group. 

In our simulation studies, we consider only positive values for $\rho^x$, $\rho^w$, $\balpha$, and $\bbeta$ for two reasons. First, a negative $\rho^x$ produces a much smaller standardized mean difference than a positive $\rho^x$. In addition, opposite signs for $\beta_1$ and $\beta_2$ could produce a total confounding close to zero because the standardized mean differences of $X_1$ and $X_2$ can be canceled out in~\eqref{eq:TC}. That is, a negative correlation between true covariates or their opposite relationship to the outcome could cancel out biases from $X_1$ and $X_2$. As a result, it is hard to find proper positive $\balpha$ values satisfying the total confounding we look for with a negative $\rho^x$ and mixed signs in elements of $\bbeta$. Second, a negative $\rho^x$ can become positive by multiplying one of the covariates by -1. Similarly, different signs between $\alpha_1$ and $\alpha_2$ can be changed into the same sign by multiplying one of the covariates by -1. To illustrate this, Figure A.1 in Appendix 2 shows that there is the same amount of bias in ATE estimates (and total confounding as well) when $\rho^x$ is negative and $\alpha_1 = \alpha_2$ and when $\rho^x$ is positive and $\alpha_1 = -\alpha_2$.


\subsubsection{Results}

\begin{figure}[htp]
    \begin{center}
        \subfigure[Bias]{\includegraphics[height=8cm]{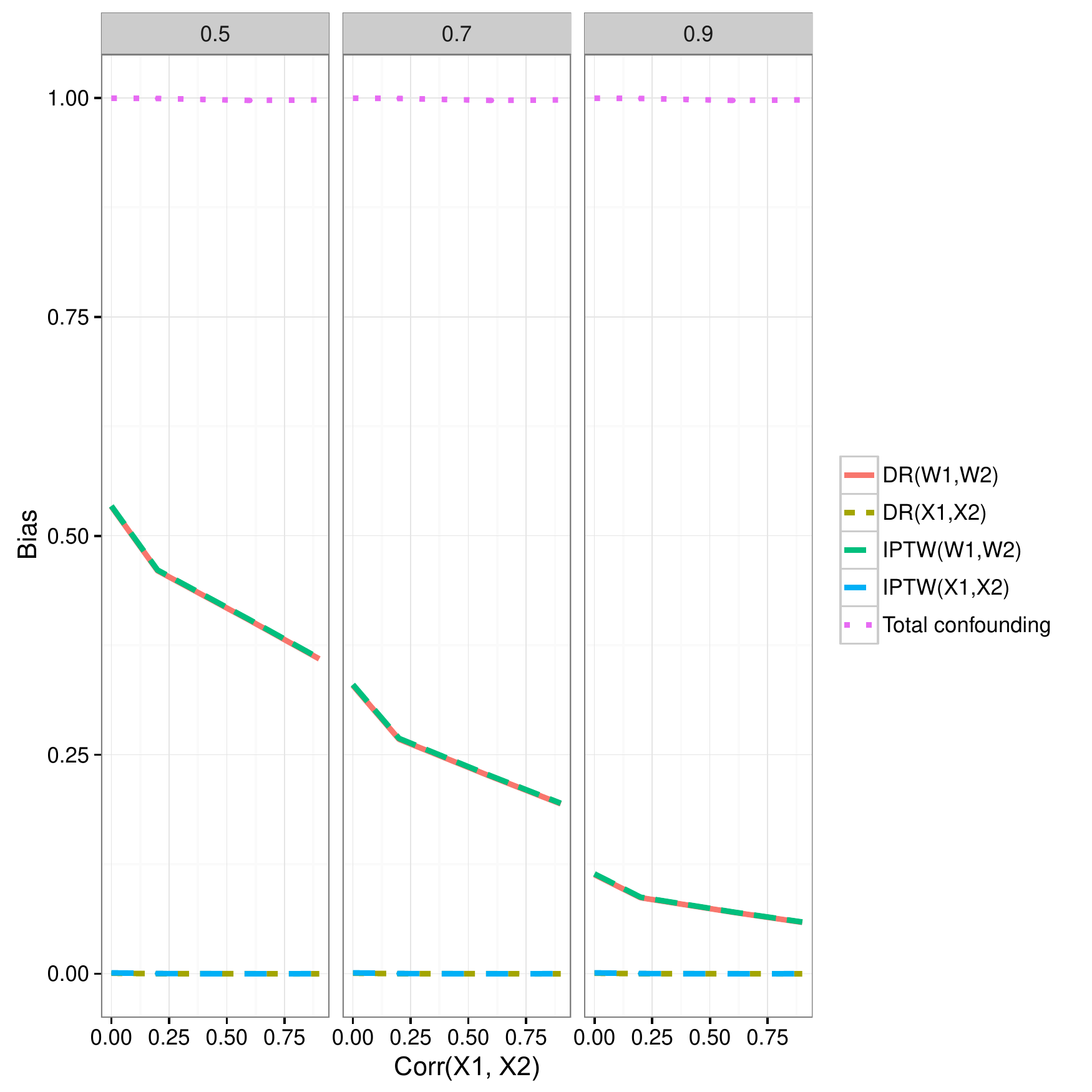}}
        \subfigure[MSE]{\includegraphics[height=8cm]{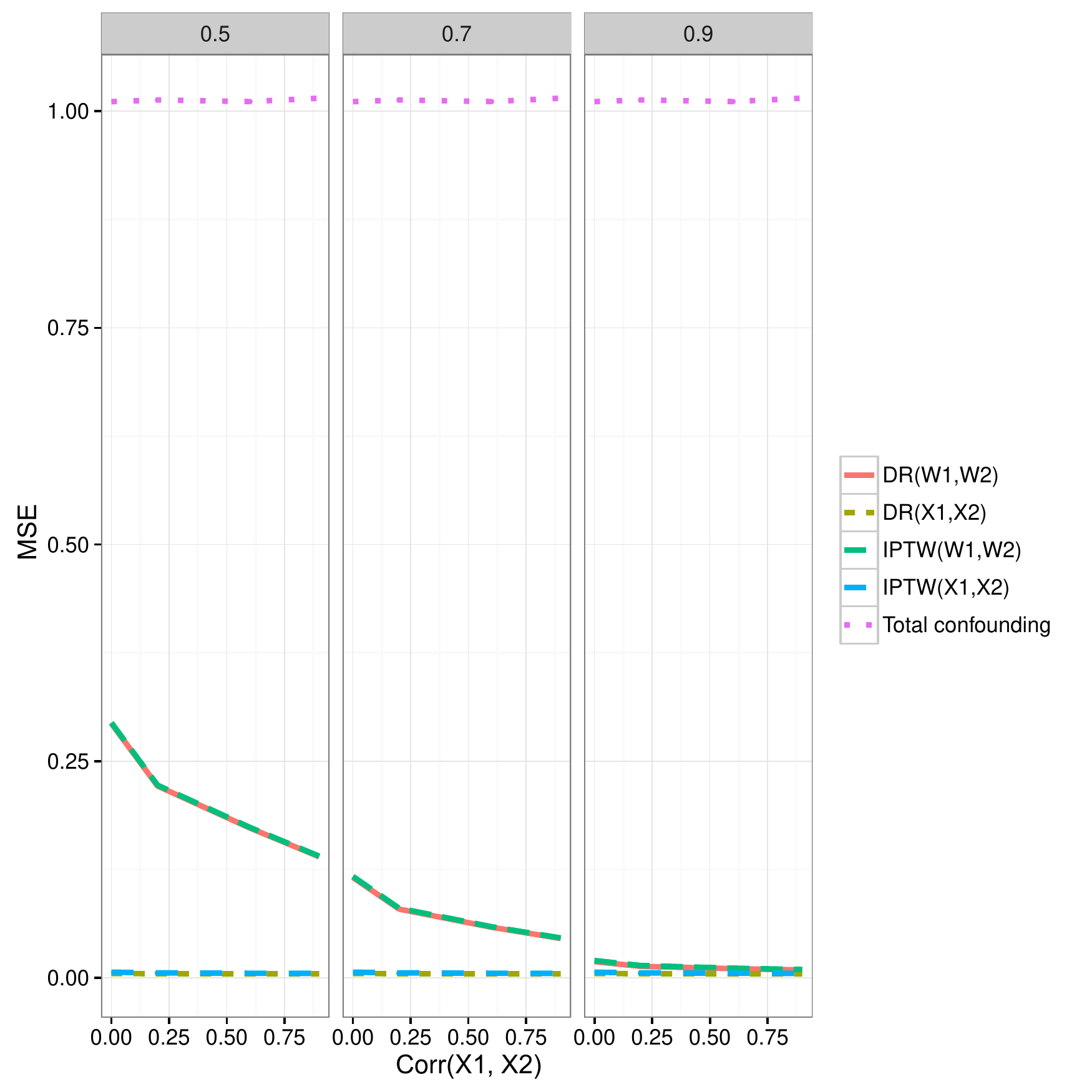}}
        \subfigure[Coverage probability]{\includegraphics[height=8cm]{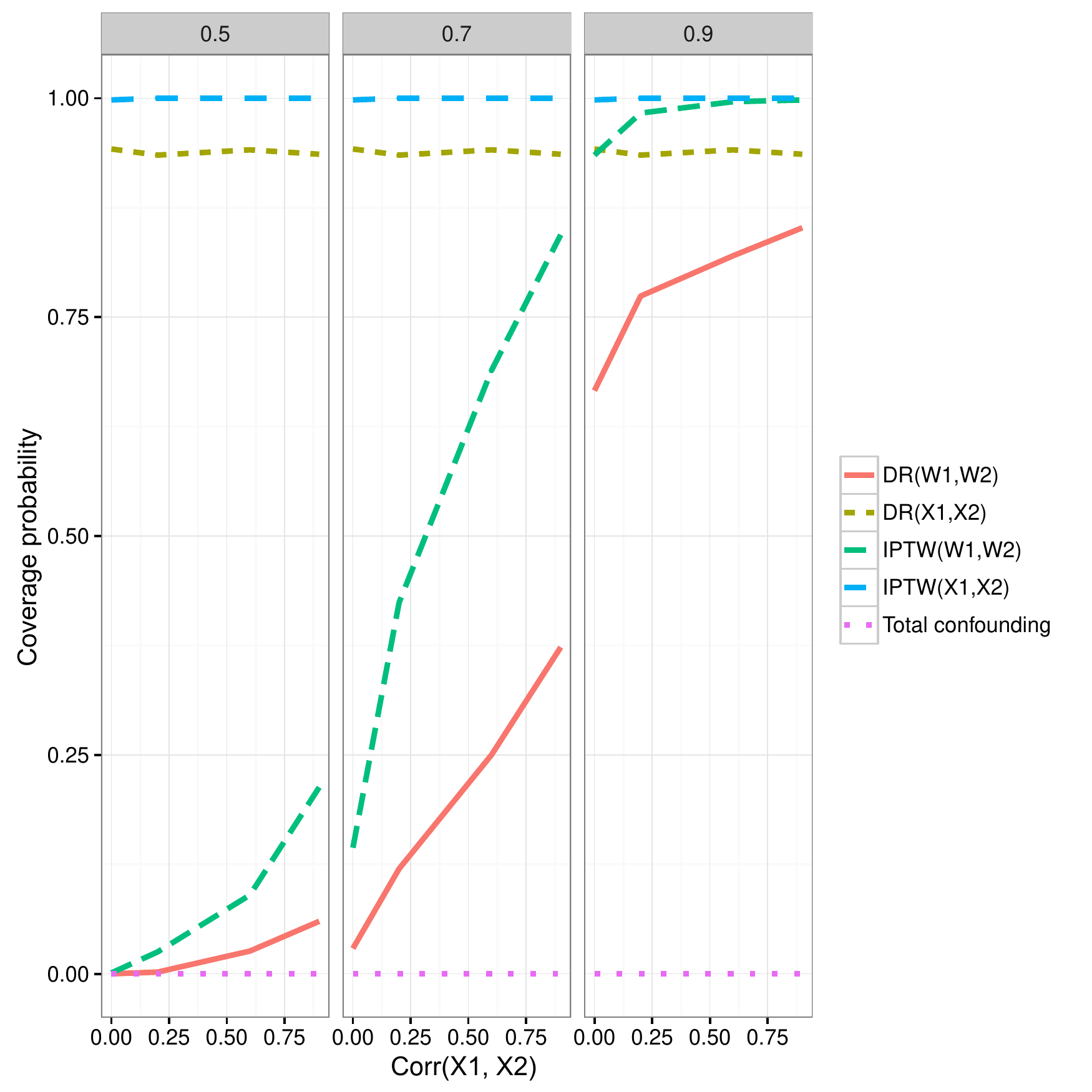}}
    \end{center}
\caption{Simulation 1 results for IPTW and DR estimates when $\rho^w=0$.}
\label{fig:sim1.1}
\end{figure}

Figure~\ref{fig:sim1.1} examines the case where there is correlation between the $\bX$'s themselves, but not between the measurement error ($\rho^w = 0$) and shows the bias, MSE, and confidence interval coverage rates of nominal 95\% intervals (called coverage probability) of four estimates: IPTW and DR with $\bX$ or $\bW$, denoted by IPTW($X_1$, $X_2$), IPTW($W_1$, $W_2$), DR($X_1$, $X_2$), and DR($W_1$, $W_2$). The three windows in each panel are for different reliability settings, 0.5, 0.7, or 0.9. In Panel (a), as expected, IPTW and DR estimators with $\bX$ produce no bias. Using $\bW$ yields biased estimates and the bias decreases as reliability becomes close to 1. Given a reliability value, biases for IPTW($W_1$, $W_2$) and DR($W_1$, $W_2$) are almost identical (hence the lines overlap), and decrease as $\rho^x$ increases. This shows that bias induced from measurement error can be reduced when the true covariates are highly correlated. The same trend is observed in terms of MSE in Panel (b). In Panel (c), the coverage probabilities of both estimates using $\bW$ increase as the correlation between $X_1$ and $X_2$ increases and reliability is close to 1. However, IPTW always provides higher coverage probability than DR. Even when we use $\bX$, the coverage probability of IPTW($X_1$, $X_2$) is 1, while that of DR($X_1$, $X_2$) is around 0.95.  

\begin{figure}[htp]
    \begin{center}
        \subfigure[Bias]{\includegraphics[height=8cm]{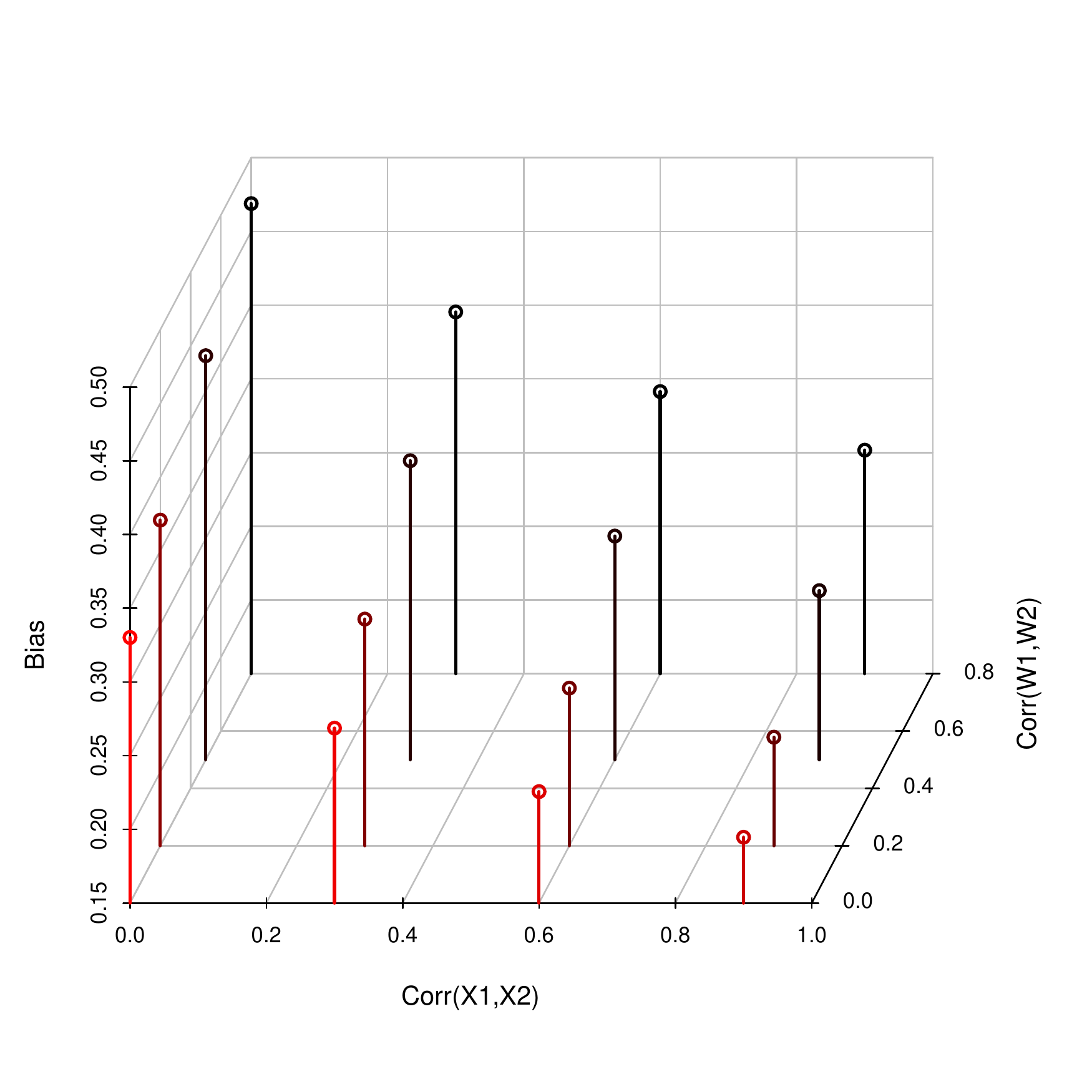}}
        \subfigure[MSE]{\includegraphics[height=8cm]{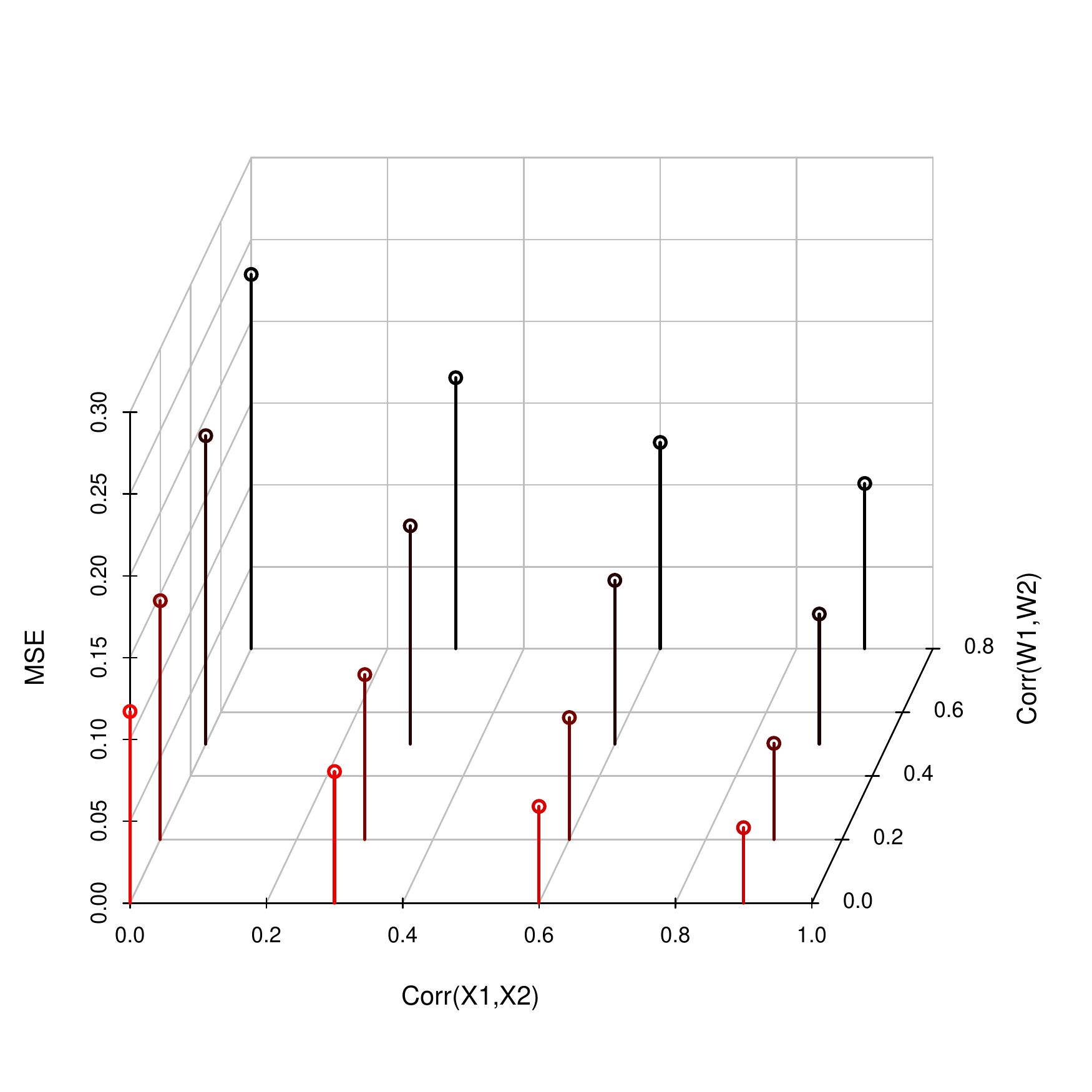}} \\
        \subfigure[Coverage probability]{\includegraphics[height=8cm]{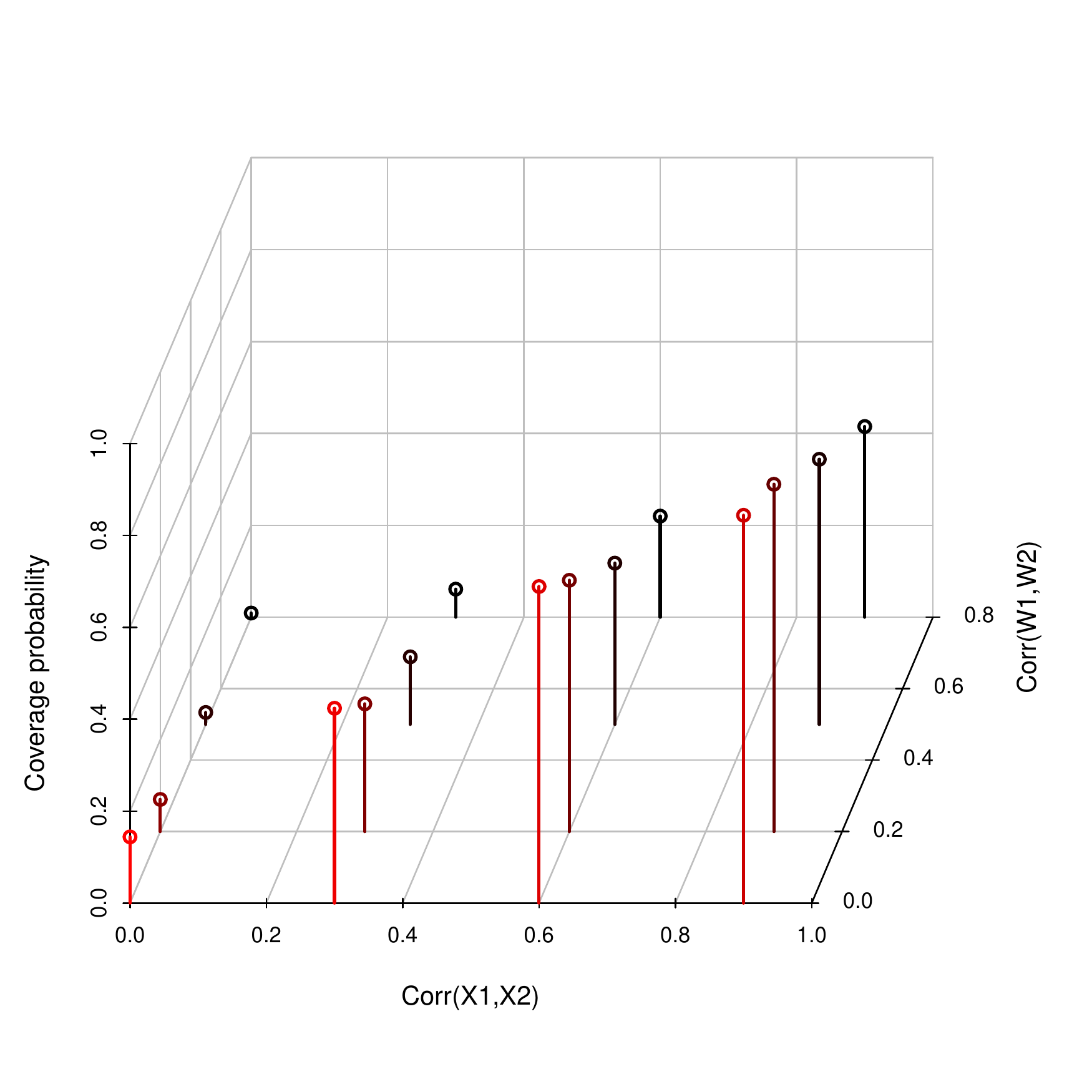}}
    \end{center}
\caption{Simulation 1 results for IPTW estimates when $\rho^w \neq 0$ and reliability=0.7.}
\label{fig:sim1.2}
\end{figure}

Now turning to the case where $X_1$ and $X_2$ are correlated and $e_1$ and $e_2$ are correlated, Figure~\ref{fig:sim1.2} shows the bias, MSE, and coverage probability of IPTW($W_1$, $W_2$). Here, we show results when reliability is 0.7. Bias and MSE increase and coverage probability decreases as the correlation between measurement errors increases, fixing the correlation between true covariates (see the y-axis from front to back). Although we found that the correlation between true covariates helps obtain a better estimate above, the correlation between measurement errors has a negative impact on effect estimation. The additional non-zero $\rho^w$ makes the propensity score weights estimated using $\bW$ deteriorate their role of balancing treatment groups with respect to the true covariates, compared to the case when $\rho^w = 0$. This results in additional bias due to the measurement error correlation and it increases as $\rho^w$ becomes larger. 


\subsection{Simulation 2: implications of measurement error in different types of covariates}

\subsubsection{Data generating mechanism and parameter setup}

In Simulation 2 we investigate how much measurement errors affect the estimates when the true covariates are not all true confounders and are actually a combination of a confounder and variables related to only the outcome or only the treatment assignment. Panel (b) of Figure~\ref{fig:simDAG} shows the data structure of interest. It includes three covariates: (1) $X_1$ related to both $A$ and $Y$ (i.e., confounder), (2) $X_2$ related to only $Y$, and (3) $X_3$ related to only $A$. Here, we assume that $X_2$ and $X_3$ are not correlated.

We generate 1000 sets of data where each dataset has 1000 observations under the following distributions:
\begin{align} \label{eq:sim2_X}
\begin{pmatrix} X_1 \\ X_2 \\ X_3 \end{pmatrix} {\sim} N\left( \begin{pmatrix} 0 \\ 0 \\ 0 \end{pmatrix}, \begin{pmatrix} 1 & \rho_{12}^x & 0 \\ & 1 & 0 \\ & & 1 \end{pmatrix} \mbox{ or } \begin{pmatrix} 1 & 0 & \rho_{13}^x \\ & 1 & 0 \\ & & 1 \end{pmatrix} \right) \\ \label{eq:sim2_A}
logit(P(A=1 \given \bX)) = \alpha_0 + \alpha_1 X_1 + \alpha_2 X_2 + \alpha_3 X_3\\ \label{eq:sim2_Y}
Y \given A, \bX {\sim} N(\beta_0 + \tau A + \beta_1 X_1 + \beta_2 X_2 + \beta_3 X_3, 1).
\end{align}
In this simulation, we consider two cases: (1) $\rho_{13}^x = 0$ and (2) $\rho_{12}^x = 0$, and the two corresponding covariance matrices are in \eqref{eq:sim2_X}. In Equation~\eqref{eq:sim2_A}, $\alpha_2$ is always set to zero because $X_2$ is not related to the treatment assignment. Similarly, $\beta_3$ in~\eqref{eq:sim2_Y} is always zero because $X_3$ is not related to the outcome. 

Now, we generate $W_p = X_p + e_p$ for $p=1, 2, 3$, where 
\begin{equation} \notag
\begin{pmatrix} e_1 \\ e_2 \\ e_3 \end{pmatrix} {\sim} N\left( \begin{pmatrix} 0 \\ 0 \\ 0 \end{pmatrix}, \begin{pmatrix} \sigma_{w_1}^2 & \sigma_{w_1}\sigma_{w_2}\rho_{12}^w & 0 \\ & \sigma_{w_2}^2 & 0 \\ & & \sigma_{w_3}^2 \end{pmatrix} \mbox{ or } \begin{pmatrix} \sigma_{w_1}^2 & 0 & \sigma_{w_1}\sigma_{w_3}\rho_{13}^w \\ & \sigma_{w_2}^2 & 0 \\ & & \sigma_{w_3}^2 \end{pmatrix} \right). 
\end{equation}
Again, we assume that $e_2$ and $e_3$ are also not correlated just like $X_2$ and $X_3$.

For the true parameter setup, we set $\sigma_{w_p}^2$ to be 1, 0.43, or 0.1 as we do in Simulation 1. When $\rho_{13}^x = 0$, we consider $\rho_{12}^x = 0.0, 0.3, 0.6,$ or 0.9, and $\rho_{12}^w = 0.0, 0.2, 0.5,$ or 0.8. Similarly, when $\rho_{12}^x = 0$, we consider $\rho_{13}^x = 0.0, 0.3, 0.6,$ or 0.9, and $\rho_{13}^w = 0.0, 0.2, 0.5,$ or 0.8. Table~\ref{table:Case2} shows the set up for $\balpha$, $\tau$, and $\bbeta$ to satisfy that the total confounding are the same across all different settings. When $\rho_{13}^x = 0$, we fix $\balpha$ and vary $\bbeta$ because varying $\balpha$ does not affect the total confounding as $X_1$ and $X_3$ are not correlated and $X_2$ has nothing to do with the treatment assignment. On the other hand, we follow fixing $\bbeta$ and varying $\balpha$ when $\rho_{12}^x = 0$ as this case is similar to Simulation 1. These parameters result in approximately half of the observations being assigned to the treated group.  

We compare 9 estimates for each IPTW and DR including different sets of covariates: (1) $\{ X_1, X_2 \}$, (2) $\{ X_1, X_2, X_3 \}$, (3) $\{ X_1, W_2, X_3 \}$, (4) $\{ X_1, X_2, W_3 \}$, (5) $\{ X_1, W_2, W_3 \}$, (6) $\{ W_1, X_2, X_3 \}$, (7) $\{ W_1, X_2, W_3 \}$, (8) $\{ W_1, W_2, X_3 \}$, and (9) $\{ W_1, W_2, W_3 \}$.

\subsubsection{Results}

For Simulation 2, we do not report DR estimates because they provide similar patterns we observed in Simulation 1: DR estimates give almost identical bias and MSE, but a bit lower coverage probabilities than IPTW estimates. In addition, we report results only when there is no correlation between measurement errors (i.e., $\rho_{12}^w$ and $\rho_{13}^w$ are zero) because correlations between measurement errors yield the same patterns we observed in Figure~\ref{fig:sim1.2} (results not shown).   

\begin{figure}[htp]
    \begin{center}
        \subfigure[Bias]{\includegraphics[height=8cm]{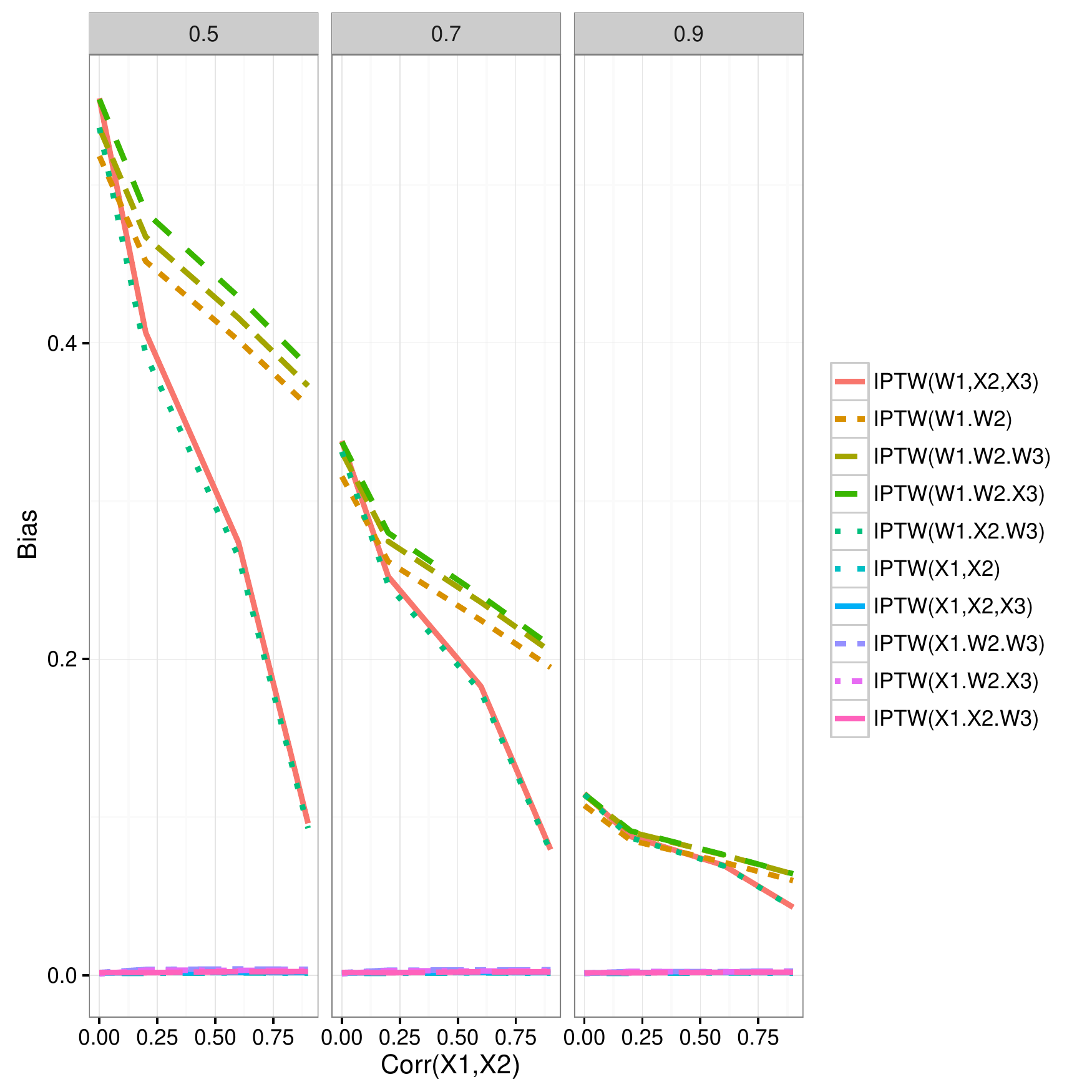}}
        \subfigure[MSE]{\includegraphics[height=8cm]{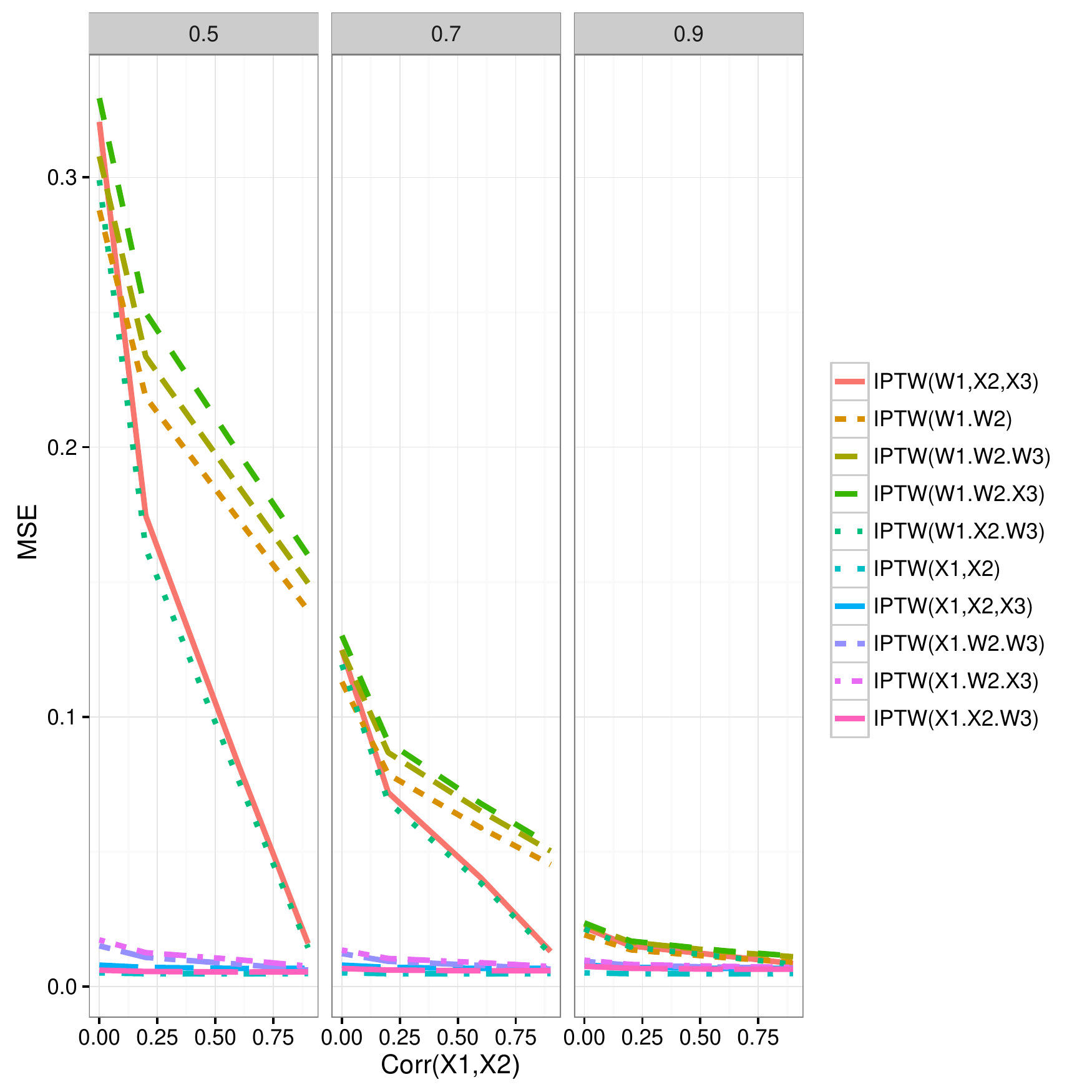}}
        \subfigure[Coverage probability]{\includegraphics[height=8cm]{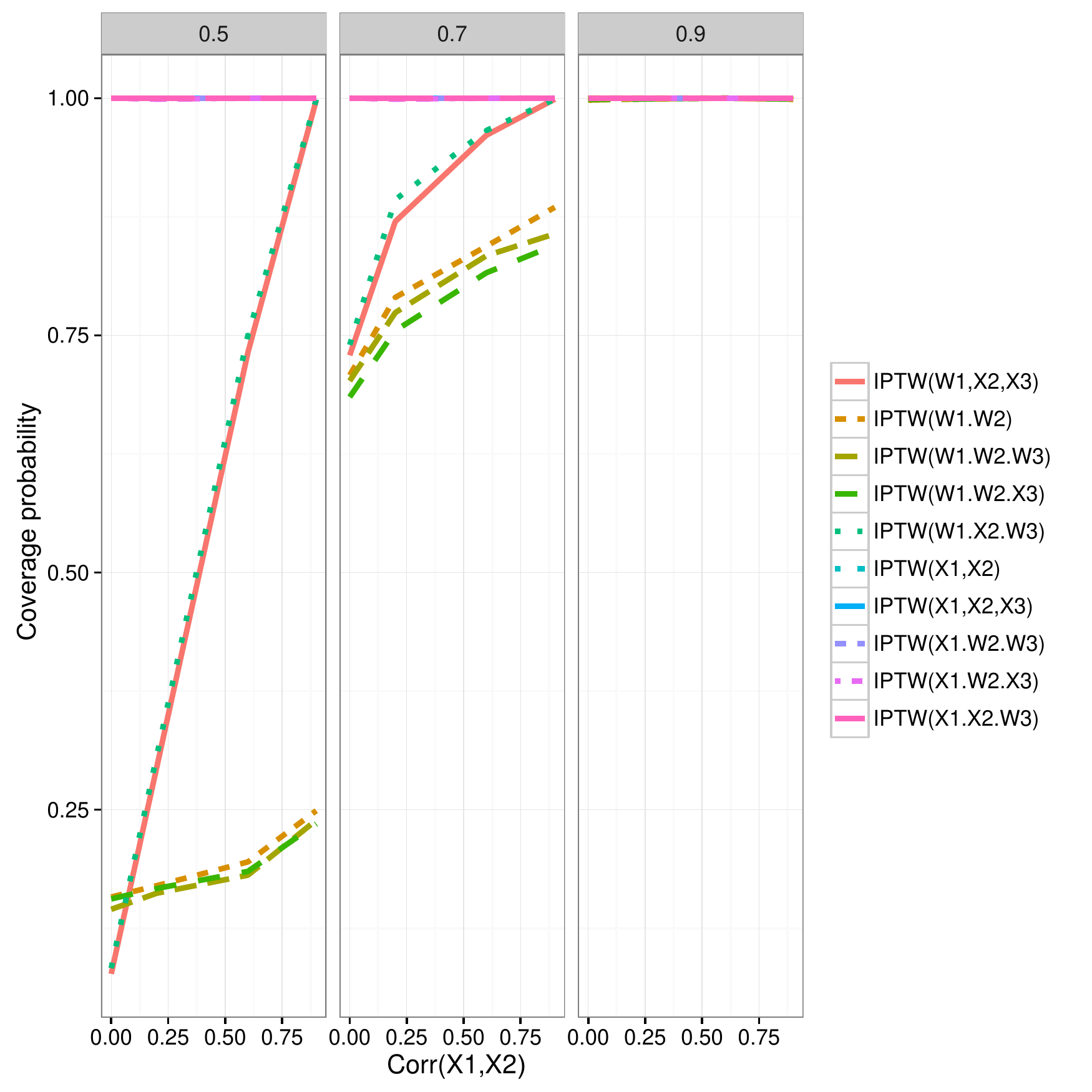}}
    \end{center}
\caption{Simulation 2 results for IPTW estimates when $\rho_{13}^x$, $\rho_{12}^w$, and $\rho_{13}^w$ are zero. Three windows are for reliability 0.5, 0.7, or 0.9.}
\label{fig:sim2.1}
\end{figure}

Figure~\ref{fig:sim2.1} displays bias, MSE, and coverage probability of 9 IPTW estimates when $\rho_{13}^x=0$, but $\rho_{12}^x \neq 0$. Three windows in each panel are for different reliability settings, 0.5, 0.7, or 0.9. The overall trend is similar as that of Simulation 1 such that bias and MSE (coverage probability) of estimates based on mismeasured covariates decrease (increases) as $X_1$ and $X_2$ become highly correlated. 

In all three panels, we categorize the 9 estimates into three groups: (1) no bias,  MSE, and almost perfect coverage, (2) relatively moderate bias, MSE, and coverage probability, and (3) high bias and MSE with low coverage probability. Note that the last two groups behave similarly when reliability is 0.9 (so many lines overlap). Group 1 contains IPTW($X_1$, $X_2$), IPTW($X_1$, $X_2$, $X_3$), IPTW($X_1$, $X_2$, $W_3$), IPTW($X_1$, $W_2$, $X_3$), and IPTW($X_1$, $W_2$, $X_3$). Group 2 contains IPTW($W_1$, $X_2$, $X_3$) and IPTW($W_1$, $X_2$, $W_3$), and Group 3 contains IPTW($W_1$, $W_2$, $X_3$), IPTW($W_1$, $W_2$, $W_3$), and IPTW($W_1$, $W_2$). There are common covariates included in each group. Group 1 (the best performing methods) always includes $X_1$, Group 2 (moderate performing methods) includes $W_1$ and $X_2$, and Group 3 (poorly performing methods) includes $W_1$ and $W_2$. As Group 1 yields no bias, as expected, this shows that including the correctly measured confounder $X_1$ is the key factor to obtain unbiased estimates. When $X_1$ is included, adding $X_3$ or $W_3$ and whether $X_2$ is measured correctly or not does not affect estimation performance. When $X_1$ is mismeasured, having $X_2$, which is associated with the outcome and correlated with $X_1$, helps obtain less biased estimates (bias and MSE in Group 2 are smaller than in Group 3). In Group 2, whether $X_3$ is measured correctly or not does not affect estimation performance. In Group 3, IPTW($W_1$, $W_2$, $X_3$) gives slightly larger bias and MSE than IPTW($W_1$, $W_2$), showing that including $X_3$ or $W_3$ neither of which is associated with the outcome and not correlated with $X_1$ does not help reduce bias and MSE. 

\begin{figure}[htp]
    \begin{center}
        \subfigure[Bias]{\includegraphics[height=8cm]{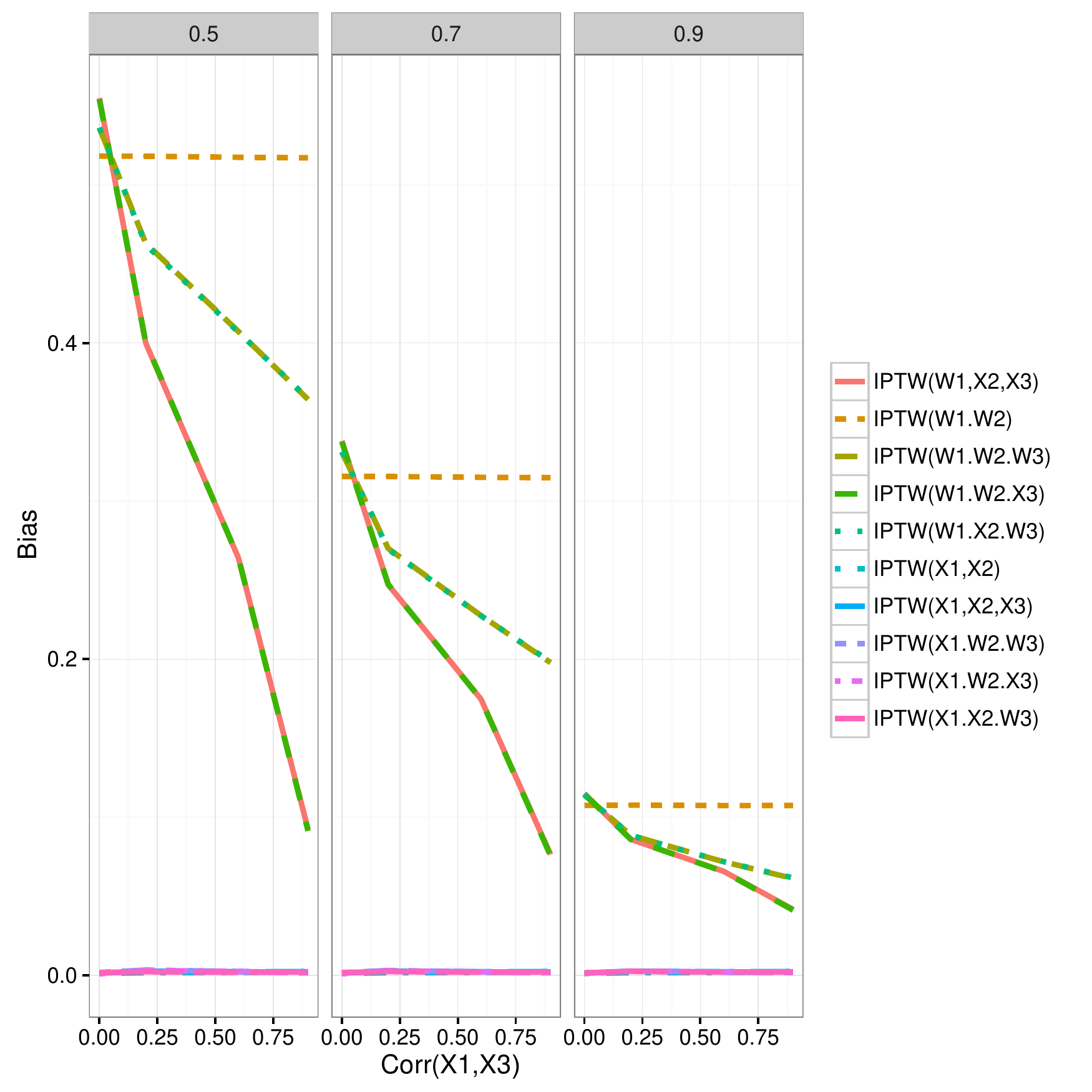}}
        \subfigure[MSE]{\includegraphics[height=8cm]{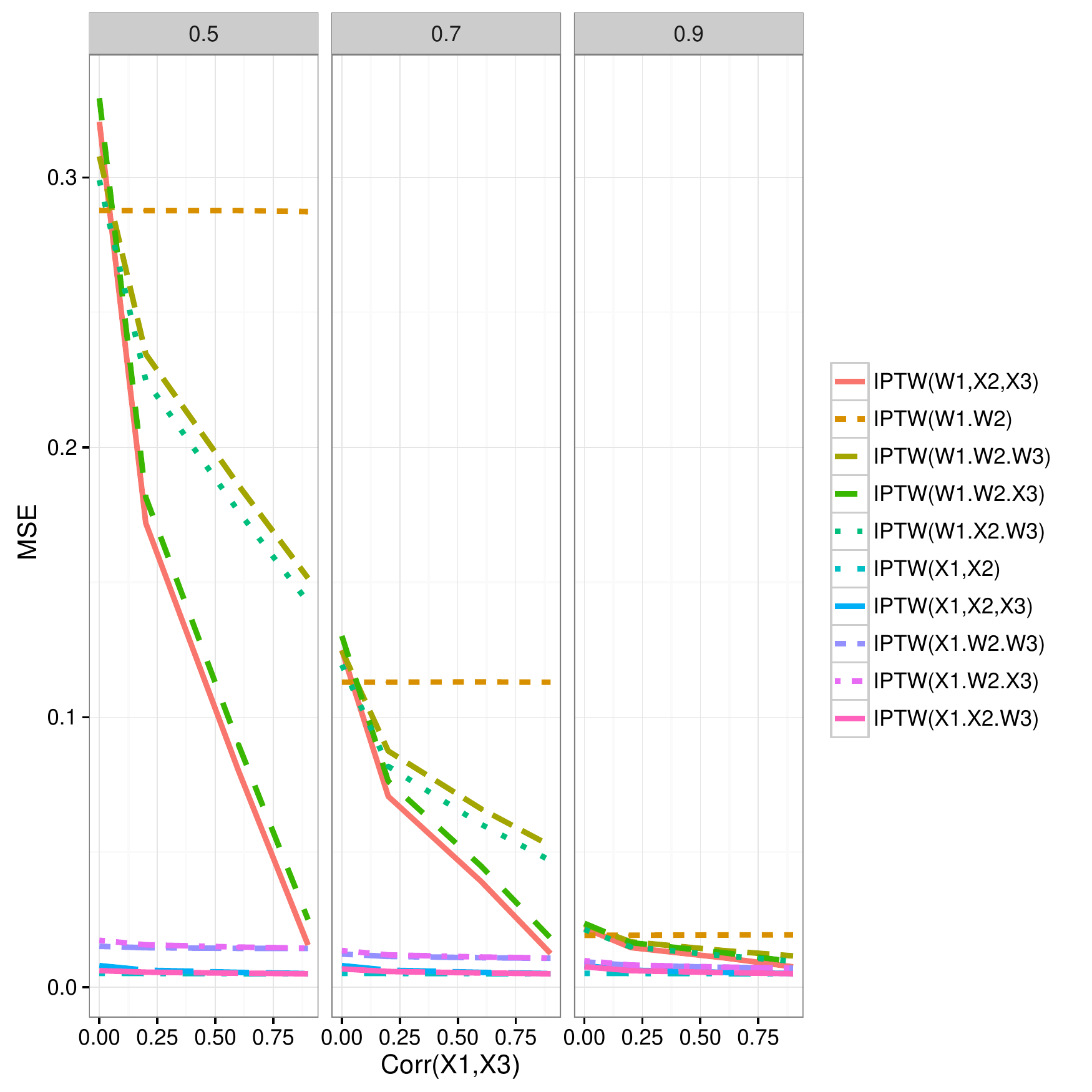}}
        \subfigure[Coverage probability]{\includegraphics[height=8cm]{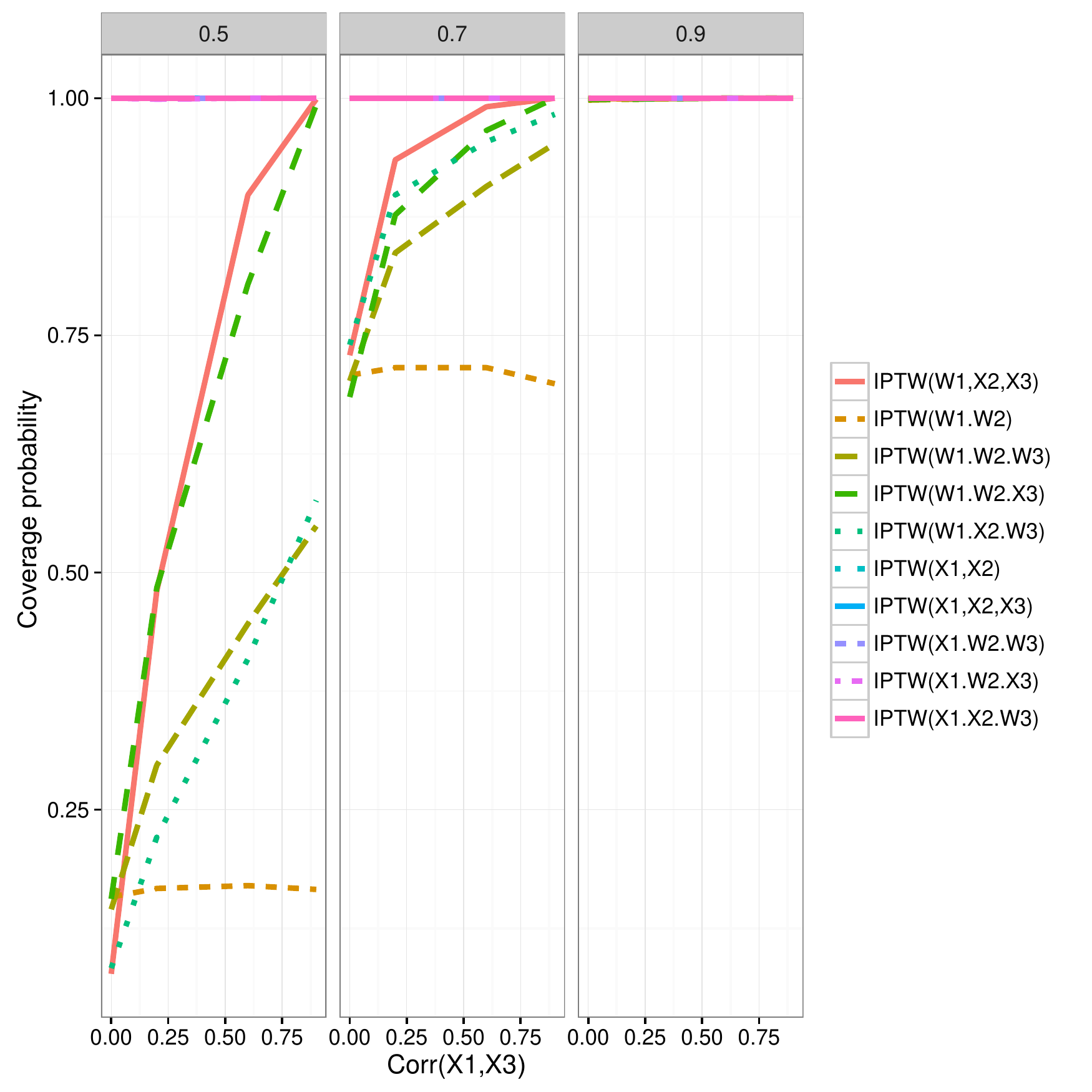}}
    \end{center}
\caption{Simulation 2 results for IPTW estimates when $\rho_{12}^x$, $\rho_{12}^w$, and $\rho_{13}^w$ are zero. Three windows are for reliability 0.5, 0.7, or 0.9.}
\label{fig:sim2.2}
\end{figure}

Figure~\ref{fig:sim2.2} shows IPTW estimates when $\rho_{12}^x$ is zero, but $\rho_{13}^x$ is not. We can now see four groups, the same three groups above and Group 4 showing the same amount of bias, MSE, and coverage probability regardless of $\rho_{13}^x$. Group 1 contains IPTW($X_1$, $X_2$), IPTW($X_1$, $X_2$, $X_3$), IPTW($X_1$, $X_2$, $W_3$), IPTW($X_1$, $W_2$, $X_3$), and IPTW($X_1$, $W_2$, $X_3$). Group 2 contains IPTW($W_1$, $X_2$, $X_3$) and IPTW($W_1$, $W_2$, $X_3$), Group 3 contains IPTW($W_1$, $X_2$, $W_3$) and IPTW($W_1$, $W_2$, $W_3$), and Group 4 contains IPTW($W_1$, $W_2$). As we observed above, all estimates in the first group commonly include $X_1$ while the other estimates include $W_1$. When $X_1$ and $X_3$ are correlated and $X_1$ is mismeasured, Group 2 provides smaller bias and MSE than Group 3, showing that it is important to include correctly measured $X_3$. Estimates in Group 4 are not affected by the correlation between $X_1$ and $X_3$ because neither $X_3$ nor $W_3$ that are associated $X_1$ is included in the estimation.   

\section{OPEN data analysis} \label{sec:data}

We now investigate the difference in estimates when using true and mismeasured covariates via a real data example. We use the OPEN study~\citep{Subar03}, which aimed to assess dietary measurement error in energy and protein intakes using two commonly used self-reported dietary instruments: the 24-hour dietary recall (24HR) and the food frequency questionnaire (FFQ). This study also measured biomarkers of these intakes using doubly labeled water and urinary nitrogen and compared measurement error between biomarkers (gold standard) and the two self-reported measurements.

Using the OPEN data, we study the relationship between smoking status (ever smoked or not) and BMI adjusting for two confounders: energy and protein intake. In this way the data example follows Figure~\ref{fig:simDAG} (a). We conduct these analyses for females (N = 223) and males (N = 261) separately, because average energy and protein intake per day are different by gender and the associations of interest may vary as well. 

We estimate the ATE using (1) biomarkers (the ``true" values), and two mismeasured versions: (2) 24HR measurements, and (3) FFQ measurements of energy and protein intakes. We  use log-transformed energy and protein intakes. Figures A.2 and A.3 in Appendix 2 show scatter plots of biomarkers and the two self-reported measurements to give a sense for the extent of the measurement error. FFQ measurements tend to under-report both energy and protein intakes as compared with the 24HR measurements. Table~\ref{table:OPENresults} shows reliability of self-reported energy and protein intake when using 24HR and FFQ by sex. For energy intake, FFQ always provides less reliable measurements than 24HR, while FFQ measurements of protein intake are less reliable than 24HR only for the male group.  

The correlations of biomarkers of energy and protein intakes are 0.47 and 0.24 for males and females, respectively. The correlations of measurement errors of self-reported energy and protein intakes are 0.65 and 0.82 for males, and 0.59 and 0.68 for females, when using 24HR and FFQ, respectively. Note that we calculate the measurement error correlations by simply taking the difference between the true value and the mismeasured value for each variable ($e = W - X$) and then calculating the correlation between the errors ($e$) for energy and protein intakes. In our example, the measurement errors are positively correlated. FFQ measurements tend to provide stronger correlations than 24HR, and the correlations are larger for males than females.    


Table~\ref{table:OPENresults} shows IPTW estimates with associated 95\% confidence intervals, standard errors, and p-values from the OPEN data analyses. Although using error-prone measurements (24HR and FFQ) does not alter significance levels, estimates based on these measurements are quite different from those estimated with the ``true" biomarkers in some cases. The ATE estimates in both male and female subgroups based on 24HR and FFQ are almost twice the estimates based on biomarkers. In addition, p-values become marginally significant for the male subgroup. Ever smoked is associated with lower BMI among females, while the opposite is observed among ever smoked males, though none of all these results are statistically significant using the standard 0.05 threshold. As we expected, there are larger differences between $\widehat{ATE}_{IPTW}$ using biomarker and self-reported measurements (i.e., more biased estimates) when using FFQ measurements because its measurement error correlation is slightly larger than the 24HR measurement error correlations for both males and females. DR estimates provide very similar results with slightly smaller standard errors (see Table A.1 in Appendix 2). 

We also reported standardized mean differences (SMDs) of the two biomarkers after weighting to examine how much propensity score weights balance exposed and unexposed groups with respect to the true confounders. For example, the SMD of the energy intake biomarker is 0.149 between ever smoked and never smoked males when estimating propensity scores using 24HR measurements, while it is 0.007 when using the true biomarkers. Overall, the balance of biomarkers between exposed and unexposed groups is poor when using self-reported measurements. 



\section{Discussion} \label{sec:Discussion}

In this paper, we investigate the impact of multiple error-prone covariates on the performance of propensity score-based causal effect estimators. Our simulation studies show that correlation between true covariates yields lower bias and MSE in ATE estimates even when mismeasured covariates are used, but that correlation between the measurement errors induces additional bias. As expected, we find that it is crucial that true confounders (variables associated with treatment and outcome) are measured without error to obtain unbiased estimates. However, when this is not feasible, including other correctly measured covariates that are correlated with the confounder can help reduce bias. This is likely related to the idea of auxiliary variables in multiple imputation~\citep{Collins01}. In addition, \cite{Rubin96} concluded that it is beneficial in terms of MSE to include covariates related to outcome, but not the treatment assignment. So covariates that are only related to the treatment should not be included in the model. However, in the context of measurement error, if a covariate only related to the treatment is correlated with confounders and the confounders are measured with error, then including the covariate helps to reduce bias and MSE.  
We illustrate the behavior of multiple error-prone covariates using real data. In the data analysis, using mismeasured covariates did not change the significance level of ATE estimates, but the point estimates sometimes changed dramatically. As such, mismeasured covariates could alter the significance level in different datasets.  

An additional innovation of this work is the introduction and use of the concept of total confounding in our simulation studies. This was crucial in our setting with correlation between covariates to isolate changes in performance due to changes in the correlation. For example, when we set the same $\balpha$ and $\bbeta$ values across all $\rho^x$ settings, we always observe increasing bias as $\rho^x$ increases. This is not a fair comparison of the role of $\rho^x$ in the measurement error context because larger $\rho^x$ creates more confounding. We want to show the beneficial role of $\rho^x$ given a fixed amount of confounding; controlling the total confounding allows us to do so. 

In our simulation studies, we only consider positive values for all parameters and the same coefficient for all covariates to find a clear trend of bias, MSE, and coverage probability. However, this is not realistic when there are many covariates in practice. We can change the sign of one covariate to have all positive coefficient when there are only two covariates, but this is also not realistic. That is, even though there are many mismeasured covariates but the resulting bias could be very small because some biases from individual covariates can be canceled out. This is a known issue in causal inference and \cite{Steiner16} show some situations where a mismeasured confounder can remove more bias than correctly measured one based on the different coefficient size for covariates.

Comparing IPTW and DR estimators, our simulation studies show that they perform similarly in terms of bias and MSE. This is expected because we do not consider model misspscification in this paper. However, IPTW tends to provide a bit wider 95\% confidence intervals than DR resulting in over coverage. Even when correctly measured covariates are included, DR provides coverage probabilities around 0.95 while IPTW provides coverage probabilities close to 1. This is because an underfitting linear model (IPTW in our case) usually leads to overestimation of the standard error of the estimate resulting in a wide confidence interval. 

As shown, measurement error in multiple covariates can have important implications for causal effect estimation. However, this topic has not been of much attention in causal inference. Future work should develop statistical methods to handle measurement error in multiple covariates. Furthermore, we need more work considering complex yet practical and realistic measurement error structures such as differential measurement error in multiple covariates.



\section*{Acknowledgment}
Drs. Hong and Stuart were supported by the National Institute of Mental Health (R01MH099010; PI to E.A.S.), and Dr. Siddique and Mr. Aaby were funded by the National Heart, Lung, and Blood Institute (R01HL127491; PI to J.S).

\newpage

\newpage

\newpage

\newpage

\newpage

\newpage

\begin{table}[htp]
  \centering
 \begin{tabular}{ccccc}
\hline
$\rho^x$  &  $\balpha$  & $\tau$ &  $\bbeta$   &  Total confounding   \\
\hline
0.0	& (0, 0.569, 0.569) & 2 & (0,1,1)	& 1 \\
0.3	& (0, 0.423, 0.423) & 2 & (0,1,1)	& 1\\
0.6	& (0, 0.336, 0.336)	& 2 & (0,1,1) & 1\\
0.9	& (0, 0.279, 0.279)	& 2 & (0,1,1) & 1\\
\hline
  \end{tabular}
  \caption{Parameter setup for Simulation 1.}\label{table:Case1}
\end{table}

\newpage

\begin{table}[htp]
  \centering
 \begin{tabular}{ccccc}
\hline
$\rho_{12}^x$  &  $\rho_{13}^x$  &  $\balpha$  &  $\bbeta$  &  Total confounding   \\
\hline
0.0	& 0 & (0, 0.575, 0, 0.575) & (0,2,2,2,0)	& 1 \\
0.3	& 0 & (0, 0.575, 0, 0.575) & (0,2,1.538,1.538,0) & 1\\
0.6	& 0 & (0, 0.575, 0, 0.575)	& (0,2,1.248,1.248,0) & 1\\
0.9	& 0 & (0, 0.575, 0, 0.575)	& (0,2,1.05,1.05,0) & 1\\
\hline
0 & 0.0 & (0, 0.575, 0, 0.575) & (0,2,2,2,0) & 1 \\
0 & 0.3 & (0, 0.427, 0, 0.427) & (0,2,2,2,0) & 1\\
0 & 0.6 & (0, 0.34, 0, 0.34)	& (0,2,2,2,0) & 1\\
0 & 0.9 & (0, 0.283, 0, 0.283)	& (0,2,2,2,0) & 1\\
\hline
  \end{tabular}
  \caption{Parameter setup for Simulation 2.}\label{table:Case2}
\end{table}

\newpage

\begin{landscape}
	\begin{table}[htp]
		\centering
		\begin{tabular}{lllccccccc}
			\hline
			Sex	&	Measurement	& \multicolumn{2}{c}{Reliability}  &	$\widehat{ATE}_{IPTW}$ (95\% CI)	&	SE	&	P-value	&  \multicolumn{2}{c}{SMD of biomarker} \\
		                 &               & Energy & Protein &   &  & & Energy & Protein \\
			\hline
			Males	&	Biomarker	&  & &	0.566	(-0.743, 1.875)	&	0.668	&	0.398 & 0.007 &	-0.007 \\
					&	24HR &	0.30 & 0.49 & 1.032	(-0.114, 2.178)	&	0.585	&	0.079 & 0.149 & -0.011	\\
					&	FFQ	  &	0.18 & 0.36 & 1.082	(-0.052, 2.216)	&	0.579	&	0.063	& 0.130  &  -0.021 \\
			Females      	&	Biomarker	& & &	-0.478	(-2.269, 1.313)	&	0.914	&	0.602	& -0.001 & 0.000 \\
					&	24HR	& 0.22 & 0.42 &	-0.787	(-2.313, 0.738)	&	0.779	&	0.313	& -0.126 & 0.043 \\
					&	FFQ	& 0.18 & 0.43 &	-0.880	(-2.389, 0.629)	&	0.77	&	0.254 &  -0.144 &	0.053 \\
			\hline
		\end{tabular}
		\caption{IPTW estimates from OPEN data analyses.}\label{table:OPENresults}
	\end{table}
\end{landscape}

\begin{thebibliography}{}


\bibitem[\protect\citeauthoryear{Collins et~al.}{2001}]{Collins01}
Collins LM, Schafer JL, and Kam CM. A comparison of inclusive and restrictive strategies in modern missing data procedures. {\it Psychological Methods} 2001;{\bf 6}: 330--351.

\bibitem[\protect\citeauthoryear{Fewell et~al.}{2007}]{Fewell07}
Fewell Z, Smith GD, and Sterne JA. The impact of residual and unmeasured confounding in epidemiologic studies: a simulation study. {\it American Journal of Epidemiology} 2007;{\bf 166}: 646--655.

\bibitem[\protect\citeauthoryear{Hong et~al.}{2016}]{Hong16}
Hong H, Rudolph K, and Stuart EA. Bayesian approach for addressing differential covariate measurement error in propensity score methods. {\it Psychometrika} 2016;doi:10.1007/s11336-016-9533-x.

\bibitem[\protect\citeauthoryear{Horvitz and Thompson}{1952}]{Horvitz52}
Horvitz DG and Thompson DJ. A generalization of sampling without replacement from a finite universe. {\it Journal of the American Statistical Association} 1952;{\bf 47}: 663–-685.


\bibitem[\protect\citeauthoryear{McCaffrey et~al.}{2013a}]{McCaffrey13a}
McCaffrey DF, Lockwood JR, and Setodji CM. Inverse probability weighting with error-prone covariates. {\it Biometrika} 2013a;ast022.

\bibitem[\protect\citeauthoryear{McCaffrey et~al.}{2013b}]{McCaffrey13b}
McCaffrey DF, Griffin BA, Almirall D, Slaughter ME, Ramchand R, and Burgette LF. A tutorial on propensity score estimation for multiple treatments using generalized
boosted models. {\it Statistics in Medicine} 2013b;{\bf 32}: 3388–-3414.

\bibitem[\protect\citeauthoryear{Pingel and Waernbaum}{2016}]{Pingel16}
Pingel R and Waernbaum I. Correlation and efficiency of propensity score-based estimators for average causal effects. {\it Communications in Statistics-Simulation and Computation} 2016;{\bf 30}: 1-21.

\bibitem[\protect\citeauthoryear{Robins et~al.}{2007}]{Robins07}
Robins J, Sued M, Lei-Gomez Q, and Rotnitzky A. Comment: Performance of double-robust estimators when ``inverse probability" weights are highly variable. {\it Statistical Science} 2007;{\bf 22}: 544--559.

\bibitem[\protect\citeauthoryear{Rosenbaum and Rubin}{1983}]{Rosenbaum83}
Rosenbaum PR and Rubin DB. The central role of the propensity score in observational studies for causal effects. {\it Biometrika} 1983;{\bf 70}: 41--55.

\bibitem[\protect\citeauthoryear{Rubin}{1974}]{Rubin74}
Rubin DB. Estimating causal effects of treatments in randomized and nonrandomized
studies. {\it Journal of educational Psychology} 1974;{\bf 66}: 688.

\bibitem[\protect\citeauthoryear{Rubin}{1980}]{Rubin80}
Rubin DB. Randomization analysis of experimental data: the Fisher randomization test comment. {\it Journal of the American Statistical Association} 1980;{\bf 75}: 591--593.

\bibitem[\protect\citeauthoryear{Rubin and Thomas}{1996}]{Rubin96}
Rubin DB and Thomas N. Matching using estimated propensity scores: relating theory to
practice. {\it Biometrics} 1996;{\bf 52}: 249--264.


\bibitem[\protect\citeauthoryear{Subar et~al.}{2003}]{Subar03}
Subar AF1, Kipnis V, Troiano RP, Midthune D, Schoeller DA, Bingham S, Sharbaugh CO, Trabulsi J, Runswick S, Ballard-Barbash R, Sunshine J, and Schatzkin A. Using intake biomarkers to evaluate the extent of dietary misreporting in a large sample of adults: the OPEN study. {\it American Journal of Epidemiology} 2003;{\bf 158}: 1--13.

\bibitem[\protect\citeauthoryear{Steiner et~al.}{2011}]{Steiner11}
Steiner PM, Cook TD, and Shadish WR. On the importance of reliable covariate measurement in selection bias adjustments using propensity scores. {\it Journal of Educational and Behavioral Statistics} 2011;{\bf 36}: 213--236.

\bibitem[\protect\citeauthoryear{Steiner and Kim}{2016}]{Steiner16}
Steiner PM and Kim Y. The mechanics of omitted variable bias: Bias amplification and cancellation of offsetting biases. {\it Journal of Causal Inference} 2016;{\bf 4}.

\bibitem[\protect\citeauthoryear{Stuart}{2010}]{Stuart10}
Stuart EA. Matching methods for causal inference: A review and a look forward. {\it Statistical science} 2010;{\bf 25}: 1.

\bibitem[\protect\citeauthoryear{St{\"u}rmer et~al.}{2005}]{Sturmer05}
St{\"u}rmer T, Schneeweiss S, Avorn J, and Glynn RJ. Adjusting effect estimates for unmeasured confounding with validation data using propensity score calibration. {\it American journal of epidemiology} 2005;{\bf 162}: 279--289.

\bibitem[\protect\citeauthoryear{Webb-Vargas  et~al.}{2015}]{Webb-Vargas15}
Webb-Vargas Y, Rudolph KE, Lenis D, Murakami P, and Stuart EA. Applying multiple imputation for external calibration to propensity score analysis. {\it Statistical Methods in Medical Research} 2015;0962280215588771.



\end{thebibliography}
\end{document}


\begin{center}
{\large \textbf{Appendix for ``Propensity score-based estimators with multiple error-prone covariates"}} \medskip

Hwanhee Hong, David A. Aaby, Juned Siddique, and Elizabeth A. Stuart
\end{center}

\section{Appendix 1. Total Confounding}

In our first simulation study, the true outcome model is 
\begin{equation} \label{eq:true_reg}
Y = \tau A + \beta_1 X_1 + \beta_2 X_2 + \epsilon.
\end{equation}
Here, we assume the intercept is zero without loss of generality. If I omit $X_1$ and $X_2$, the model is rewritten as 
\begin{equation} \label{eq:omitted_reg}
Y = \tau^* A + \epsilon^*.
\end{equation}
The estimated $\tau^*$ under~\eqref{eq:omitted_reg} is 
\begin{equation} \label{eq:beta_hat_star}
\hat{\tau^*} = (A^T A)^{-1} A^T Y.
\end{equation}
To calculate the bias of $\hat{\tau^*}$ under the true outcome model~\eqref{eq:true_reg}, let's first plug in~\eqref{eq:true_reg} to~\eqref{eq:beta_hat_star}. 
\begin{align} \notag
\hat{\tau^*} = & (A^T A)^{-1} A^T Y \\ \notag
                = & (A^T A)^{-1} A^T (\tau A + \beta_1 X_1 + \beta_2 X_2 + \epsilon) \\ \notag
                = & (A^T A)^{-1} A^T A \tau + (A^T A)^{-1} A^T X_1 \beta_1 + (A^T A)^{-1} A^T X_2 \beta_2 + (A^T A)^{-1} A^T \epsilon
\end{align}
Then,
\begin{align} \notag
E(\hat{\tau^*} \given A) = & \tau + (A^T A)^{-1} E(A^T X_1 \given A) \beta_1 + (A^T A)^{-1} E(A^T X_2 \given A) \beta_2 \\ \notag
= & \tau + bias.        
\end{align}
We define this bias as our total confounding. We can rewrite $(A^T A)^{-1} E(A^T X_1 \given A) = E(X_1 \given A=1)-E(X_1 \given A=0)$ because the left hand side is the estimated coefficient of a regression model of $X_1 {\sim} A$. Then, the total confounding can be rewritten as
\begin{equation} \notag
\beta_1[E(X_1 \given A=1)-E(X_1 \given A=0)] + \beta_2 [E(X_2 \given A=1)-E(X_2 \given A=0)].
\end{equation}

\section{Appendix 2. Additional results}

\begin{figure}[htp]
    \begin{center}
        \includegraphics[height=17cm]{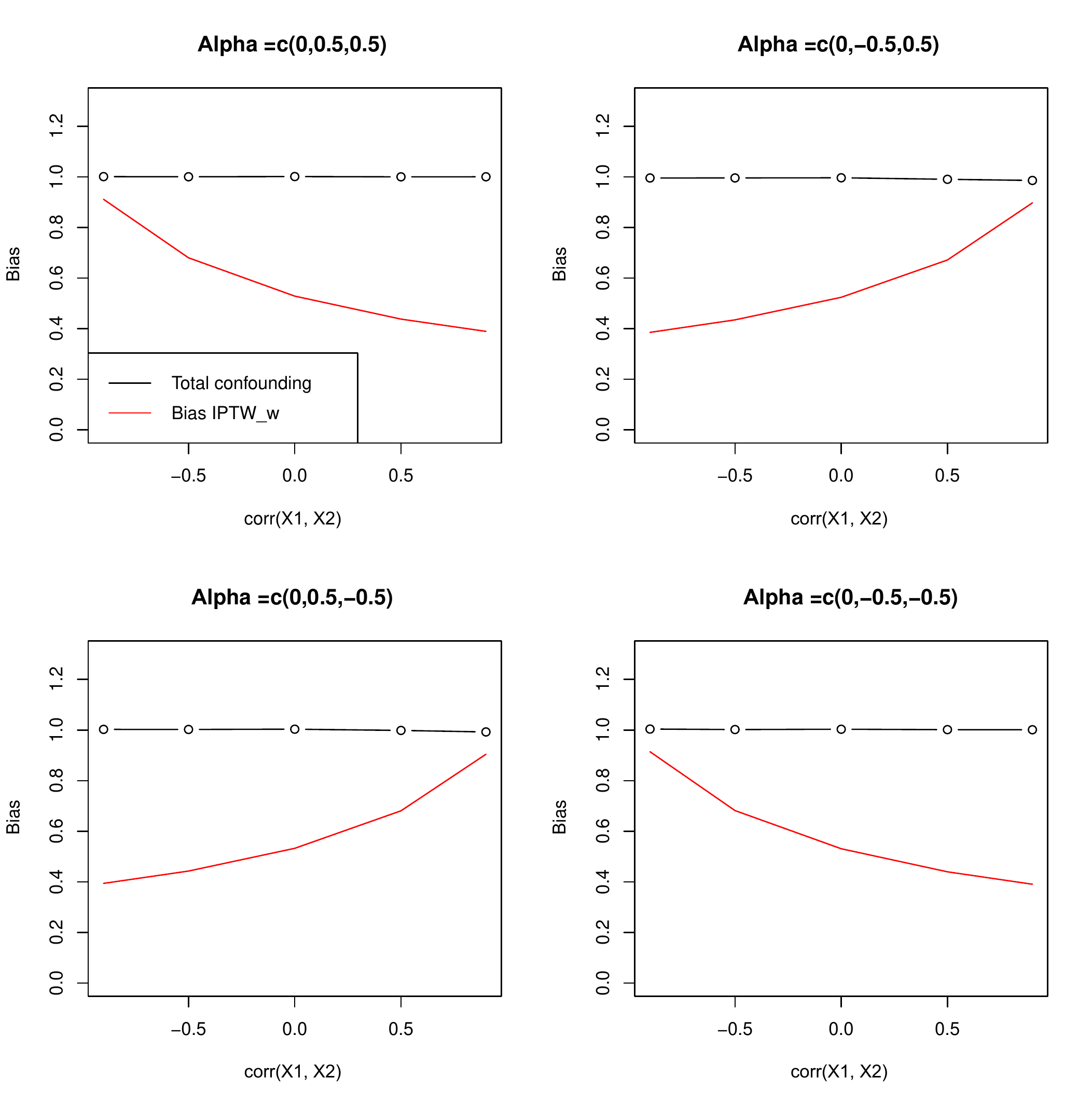}
    \end{center}
\caption{Bias of the IPTW estimator in Simulation 1 when data are generated using negative values for $\rho^x$ and $\balpha$. We vary $\bbeta$ for this plots to have the same total confounding across all settings.} \label{table:negative}
\end{figure}

\begin{figure}[htp]
    \begin{center}
        \subfigure[Male]{\includegraphics[height=9cm]{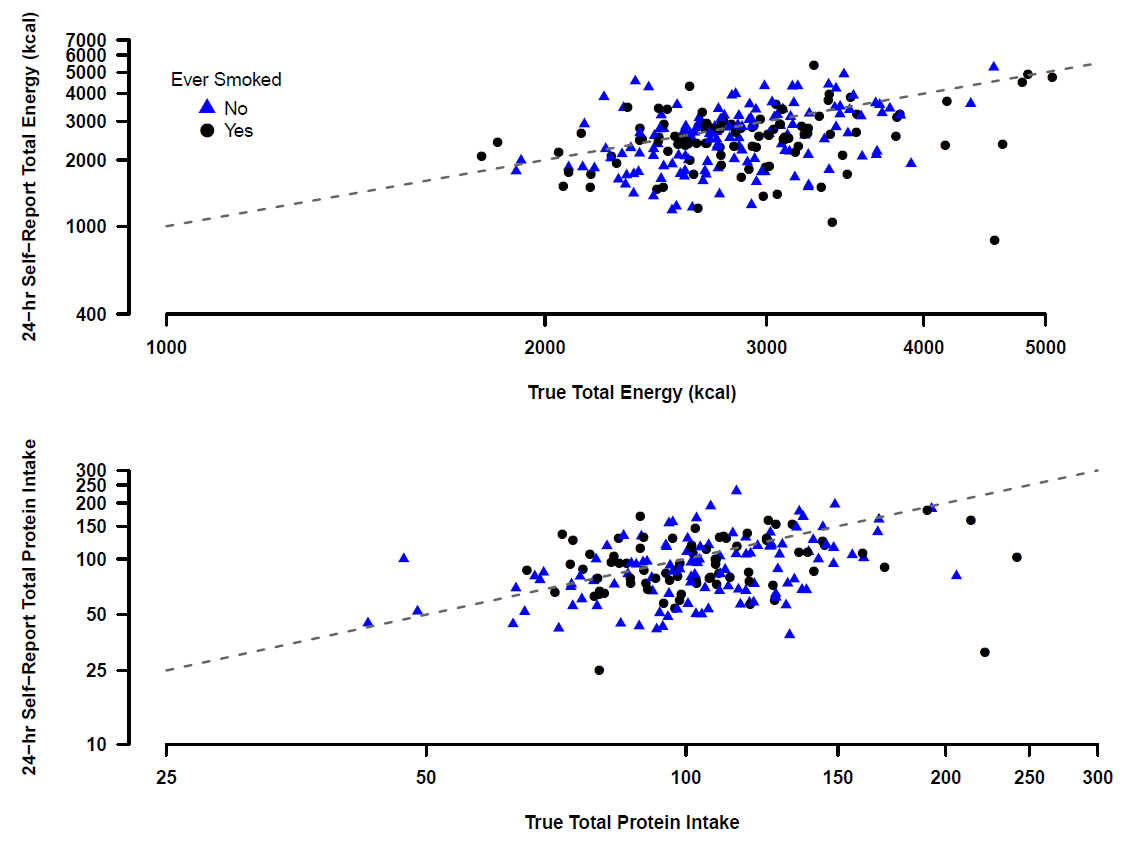}}
        \subfigure[Female]{\includegraphics[height=9cm]{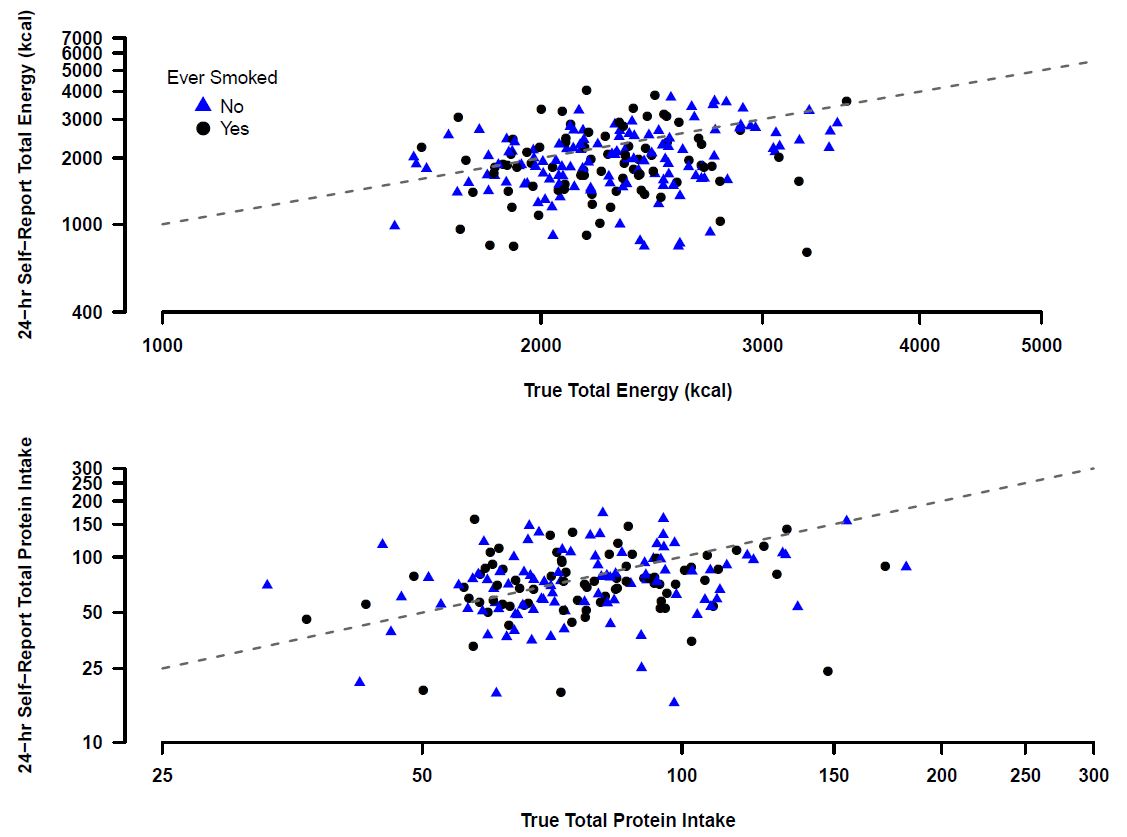}}
    \end{center}
\caption{Scatter plots of biomarker and 24HR measurement of energy and protein intakes with 45-degree dashed lines among (a) males and (b) females.}
\label{fig:OPEN}
\end{figure}

\begin{figure}[htp]
    \begin{center}
        \subfigure[Male]{\includegraphics[height=9cm]{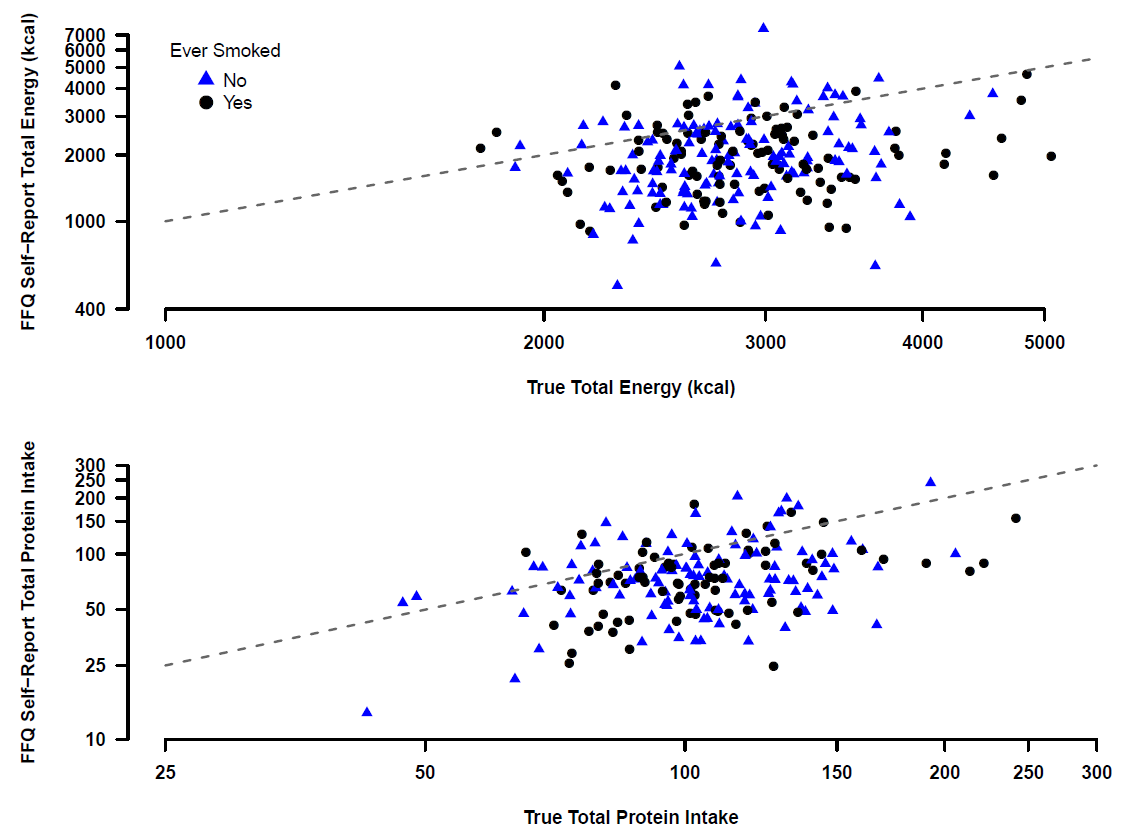}}
        \subfigure[Female]{\includegraphics[height=9cm]{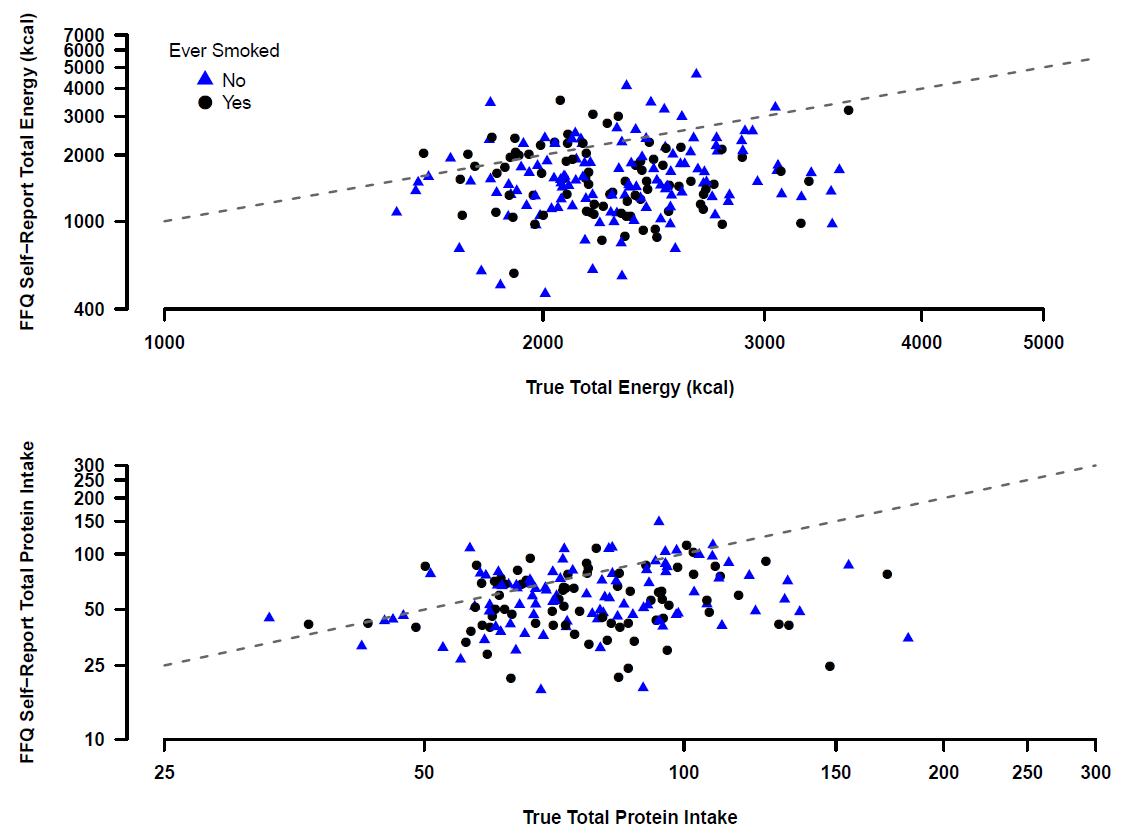}}
    \end{center}
\caption{Scatter plots of biomarker and FFQ measurement of energy and protein intakes with 45-degree dashed lines among (a) males and (b) females.}
\label{fig:OPEN}
\end{figure}

\begin{table}[h]
	\centering
	\begin{tabular}{llccc}
		\hline
			Sex	&	Measurement	&	$\widehat{ATE}_{DR}$ (95\% CI)	&	SE	&	P-value	\\
		\hline
			Males	&	Biomarker	&	0.553	(-0.538, 1.645)	&	0.557	&	0.322 	\\
				&	24HR	&	1.03	(-0.106, 2.166)	&	0.58	&	0.077 	\\
				&	FFQ	&	1.084	(-0.038, 2.207)	&	0.573	&	0.059 	\\
			Females	&	Biomarker	&	-0.476	(-1.951, 0.999)	&	0.752	&	0.528 	\\
				&	24HR	&	-0.787	(-2.311, 0.737)	&	0.777	&	0.313 	\\
				&	FFQ	&	-0.872	(-2.354, 0.611)	&	0.756	&	0.25 	\\
		\hline
	\end{tabular}
	\caption{DR estimates from OPEN data analyses. Reliabilities and SMDs of biomarkers are omitted because they are the same as in Table 3 in the main manuscript.}\label{table:OPENresults}
\end{table}